 \documentclass[preprint2]{aastex}

\slugcomment{Draft, to be submitted to the Astronomical Journal}

\shorttitle{Fading AGN properties and history}
\shortauthors{Keel et al.}

\begin{document}


\title{Fading AGN Candidates: AGN Histories and Outflow Signatures\footnote{Based on observations with the 
NASA/ESA {\it Hubble Space Telescope} obtained at the Space Telescope Science Institute, which is operated
by the Association of Universities for Research in Astronomy, Inc.,
under NASA contract No. NAS5-26555.}}


\author{William C. Keel \altaffilmark{1,2}, Chris J. Lintott\altaffilmark{3}, 
W. Peter Maksym\altaffilmark{1,4},
Vardha N. Bennert\altaffilmark{5}, S. Drew Chojnowski\altaffilmark{6},
Alexei Moiseev\altaffilmark{7}, Aleksandrina Smirnova\altaffilmark{7},
Kevin Schawinski \altaffilmark{8}, Lia F. Sartori\altaffilmark{8}, 
C. Megan Urry\altaffilmark{9}, Anna Pancoast\altaffilmark{4,10},
Mischa Schirmer\altaffilmark{11},
Bryan Scott\altaffilmark{5}, Charles Showley\altaffilmark{5}, and Kelsi Flatland\altaffilmark{5}}







\altaffiltext{1}{Department of Physics and Astronomy, University of Alabama, Box 870324, Tuscaloosa, AL 35487}
\altaffiltext{2}{wkeel@ua.edu, Twitter: @NGC3314}
\altaffiltext{3}{Astrophysics, Oxford University; and Adler Planetarium, 1300 S. Lakeshore Drive, Chicago, IL 60605}
\altaffiltext{4}{Current address: Center for Astrophysics, 60 Garden St., Cambridge, MA 02138}
\altaffiltext{5}{Physics Department, California Polytechnic State University,
San Luis Obispo, CA 93407}
\altaffiltext{6}{Department of Astronomy, New Mexico State University, P. O. Box 30001, MSC 4500, Las Cruces, New Mexico 88003-8001}
\altaffiltext{7}{Special Astrophysical Observatory, Russian Academy of Sciences, Nizhny Arkhyz, Russia 369167}
\altaffiltext{8}{Institute for Astronomy, ETH Z\"urich, Wolfgang-Pauli-Stra\ss e 27,
CH-8093 Zurich, Switzerland}
\altaffiltext{9}{Department of Physics, Yale University,
P.O. Box 208120, New Haven, CT 06520-8120}
\altaffiltext{10}{Einstein Fellow}
\altaffiltext{11}{Gemini Observatory, La Serena, Chile}


\begin{abstract}
We consider the energy budgets and radiative history of eight fading AGN, identified from an energy shortfall
between the requirements to ionize very extended (radius $>10$ kpc)
ionized clouds and the luminosity of the nucleus as we view it directly. All show evidence of significant fading on 
$\approx 50,000$-year timescales. 
We explore the use of minimum ionizing luminosity $Q_{ion}$ derived from photoionization balance in the brightest pixels in H$\alpha$ at each projected radius.  Tests using presumably constant Palomar-Green (PG) QSOs, and one of our targets with detailed photoionization modeling, suggest that we can derive useful histories of individual AGN, with the caveat that the minimum ionizing luminosity is always an underestimate and subject to uncertainties about fine structure in the ionized material.  These consistency tests suggest that the degree of underestimation from the upper envelope 
of reconstructed   $Q_{ion}$ values is roughly constant for a given object and therefore does not prevent such derivation.
The AGN in our sample show a range
of behaviors, with rapid drops and standstills; the common feature is a rapid drop in the last
$\approx 2 \times 10^4$ years before the direct view of the nucleus. The $e$-folding timescales for ionizing luminosity 
are mostly in the thousands of years, with a few episodes as short as 400 years. 
In the limit of largely obscured AGN, we find additional evidence for fading from the shortfall between even the lower limits from recombination balance and the maximum luminosities 
derived from from infrared fluxes.
We compare these long-term light curves,
and the occurrence of these fading objects among all optically identified AGN, to simulations of
AGN accretion; the strongest variations on these timespans are seen in models with strong and 
local (parsec-scale) feedback. We present Gemini integral-field optical spectroscopy, which
shows a very limited role for outflows in these ionized structures. While rings and loops of
emission, morphologically suggestive of outflow, are common, their kinematic structure
shows some to be in regular rotation. UGC 7342 exhibits local
signatures of outflows $< 300$ km s$^{-1}$, largely associated with very diffuse emission, 
and possibly entraining gas in one of the 
clouds seen in HST images. Only in the Teacup AGN do we see outflow signatures
of order 1000 km s$^{-1}$. In contrast to the extended emission regions around many radio-loud
AGN, the clouds around these fading AGN consist largely of tidal debris being externally 
illuminated but not displaced by AGN outflows.
\end{abstract}




\newpage
\section{Introduction}

It has long been known that some active galactic nuclei (AGN) are accompanied by extended emission-line regions (EELRs), zones of ionized gas spanning galaxy scales or even
larger. Such regions can trace the geometry of ionizing radiation escaping the AGN and 
host galaxy, and at least implicitly give hints to the luminosity history of the AGN. EELRs
occur around Seyfert nuclei, QSOs, and radio galaxies. 

A very luminous and extensive EELR was found in the course of the Galaxy Zoo project \citep{cjl2008}
near the spiral galaxy IC 2497. Known after its discoverer as Hanny's Voorwerp,
this object shows high-ionization emission lines with ratios essentially identical
to the narrow-line region of an AGN, spanning a projected range from 15-35 kpc
from the galaxy nucleus, which fails by at least 2 orders of magnitude in bolometric
output to match the ionization requirements of the cloud
(\citealt{cjl2009}, \citealt{HSTHV}). The cloud's electron temperature, narrow line widths,
and quiescent velocity field indicate photoionization rather than shocks as the energy
source. H I mapping by \cite{Jozsa} shows this to be the ionized part of a 300-kpc trail
of otherwise neutral gas, suggesting a strong galaxy interaction many Gyr ago. The energy mismatch 
between the nucleus of IC 2497 and the ionization requirements in Hanny's Voorwerp 
indicates that the nucleus faded from a luminosity associated with QSOs to a modest
Seyfert or LINER level within $10^5$ years.

Starting with the unusual morphology and SDSS $gri$ colors of Hanny's Voorwerp, a targeted
search by Galaxy Zoo volunteers found 19 additional EELRs, including hitherto unknown
examples, of which 8 showed energy shortfalls similar to that in IC 2497 \citep{mnras2012}.
This sample was selected in a homogeneous way, requiring detection of emission features 
showing AGN-like line
ratios more than 10 kpc in projection from the nucleus. We have pursued detailed study of 
this subset of 8 (plus IC 2497/Hanny's Voorwerp) to learn more about AGN variations on the otherwise inaccessible range of time scales $10^4-10^5$ years.
Analogous cases of extended ionized gas near energetically inadequate AGN have been
reported at higher redshift and luminosity by \cite{Schirmer} and \cite{Schirmer2016}, their ``Green Bean" systems, and
at low luminosity in the local Universe in the merger remnant NGC 7252 \citep{Schweizer}
and as one interpretation of an off-nuclear ionized region in the spiral galaxy NGC 3621 \citep{menezes}.
	
Interest in the history of accretion onto nuclear black holes has also been renewed by
evidence for an eventful history of outbursts within $10^6$ years by  the currently quiescent 
black hole at the
Galactic Center, from X-ray echoes \citep{Muno},
the ionization structure of the Magellanic Stream \citep{magstream},
and possibly in the ``Fermi bubbles" ({\citealt{zkn2011}, \citealt{GuoMathews}}).

The variations in AGN luminosity we infer from these objects, spanning up to $10^5$-year scales,
connect implicitly with the variations on timescales we can observe
directly.  ``Changing-look" AGN can have broad-line emission 
essentially vanish over $\approx 10$ years in the emitted frame ({\citealt{MacLeod},
\citealt{Ruan}, \citealt{Runnoe}, \citealt{Runco}), while
the narrow-line region is often so large that light-travel-time smearing leaves its emission 
nearly constant;
similar interpretations were discussed in the context of spectral variations in Seyfert nuclei at
least as early as \cite{PenstonPerez}.
The temporal spectrum of variations in AGN output 
clearly spans many orders of magnitude, and affects our understanding not only 
of their physical structure but the demographics of accretion in the galaxy 
population. Seeing dim episodes so long that they appear in surrounding ionized gas when smeared by the recombination timescales of thousands of years, as well as geometric projection factors,
means that the AGN either stay in a very low state for these long times, or that 
excursions to a high state fill only a small fraction of the time. Understanding this history is key to understanding the broad
idea of duty cycles in AGN accretion; X-ray surveys in particular suggest
a duty cycle connected to the Eddington limit and evolving with 
cosmic time (e.g., \citealt{Shankar}), but not how many
episodes of what durations are involved.

In a similar sense, radio observations suggest very episodic
production of jets, a major form of kinetic energy output from accretion.
The identification of radio jets and a circumnuclear outflow in IC 2497
(\citealt{Jozsa}, \citealt{Rampadarath}, \citealt{HSTHV}) led us to speculate that some of the extreme variability needed to explain the
ionization of Hanny's Voorwerp might come not solely from a drop in the 
accretion-driven luminosity of the AGN in IC 2497, but from some of the ``missing"
luminosity switching to a kinetic mode. This idea is supported by a ``bubble"
of hot gas around the fading AGN of IC 2497 seen with {\it Chandra}
\citep{Sartori}.

All these factors motivated more detailed study of our fading-AGN candidates
with a variety of techniques.
Paper I \citep{paper1}
presented results of HST imaging concerning the host galaxies and
origin of the ionized gas. The hosts are bulge-dominated, and every one shows
signatures of ongoing or past galaxy interactions, 1.5-3 Gyr ago for systems with
favorable geometry to estimate this timescale.
The extended gas has modestly subsolar
metallicity, and is kinematically rotation-dominated; outflows have only a very
localized role, in contrast to most QSO EELRs \citep{SFC2006}. These are
externally illuminated tidal debris; merger remnants are particularly good 
environments for H I to occur far from the nucleus and out of any host disk plane,
acting as a screen to show the escaping ionizing radiation. This appears to be
such a strong selection factor that we do not know whether 
high-amplitude variability on $10^4$-year scales is confined to merger aftermaths
or not.

In this paper, we describe additional data, including Gemini integral-field spectroscopy,
and present implications of the entire data set for the properties and history of 
these AGN. We use the peak surface brightness in recombination lines at various
projected radii to reconstruct the luminosity history of the AGN, compare structures
seen in the emission-line images and line ratios to address the role of ionization cones,
and examine the occurrence of outflows near the nuclei to trace the role and extent of
feedback near the epochs of radiative dimming.


In evaluating sizes and luminosities, we compute distances using a Hubble constant 
73 km s$^{-1}$ Mpc$^{-1}$.

\section{Observations}

\subsection{Galaxy sample}

As given by \cite{paper1}, the sample of AGN we consider includes the eight systems 
from the Galaxy Zoo survey of giant ionized clouds \citep{mnras2012} where a 
simple energy-budget estimate including the ionization requirements of the most distant
emission line region, compared to the observed bolometric luminosity of the nuclei, showed
a shortfall greater than about a  factor 5. These include NGC 5252 with its well-documented
set of ionization cones (\citealt{PrietoFreudling}, \citealt{morse}), and the large-scale double radio source 
hosts NGC 5972 \citep{VCV2001}
and Mkn 1498 \citep{Huub}. Additional objects, some showing larger and more luminous EELRs, were newly found in this survey. This work deals with the $\approx 40$\% of the Galaxy Zoo EELR objects showing evidence for AGN fading; the remainder have energy budgets indicating
roughly constant luminosity in some cases with substantial obscuration along our line of sight.
Overall properties of these galaxies are listed in Table \ref{tbl-galaxies}, including IC 2497 (host
galaxy of Hanny's Voorwerp) for reference.

\subsection{HST imaging}

Central to our analysis is a set of emission-line images of our fading-AGN candidates obtained with the Hubble Space Telescope (HST) and narrowband filters
in the Wide Field Camera 3 (WFC3) or 
tunable ramp filters on the Advanced Camera for Surveys (ACS), using archival WFPC2 data 
for NGC 5252. These data and our processing are described in detail by \cite{paper1}.
For brevity, we truncate the SDSS designations of two systems without more common
catalog names, SDSS J151004.01+074037.1 and SDSS J220141.64+115124.3.

\begin{deluxetable}{llccc}
\tablecaption{Galaxy properties  \label{tbl-galaxies}}
\tablewidth{0pt}
\tablehead{
\colhead{Galaxy} & \colhead{SDSS Name} & \colhead{Type} & \colhead{$z$}  &  \colhead{Scale, kpc/\arcsec}
}
\startdata
IC 2497 & SDSS J094104.11+344358.4 & 2 & 0.0502 & 0.96 \cr 
Mkn 1498 &  SDSS J162804.06+514631.4  & 1.9 & 0.0547 	& 1.11\cr
NGC 5252 & SDSS J133815.86+043233.3  & 1.5 & 0.0228 & 0.50  \cr
NGC 5972 & SDSS J153854.16+170134.2 & 2 & 0.0297 & 0.63 \cr
SDSS 1510+07 &  SDSS J151004.01+074037.1 & 2 & 0.0458 &  0.89  \cr
SDSS 2201+11 & SDSS J220141.64+115124.3 & 2 & 0.0296 & 0.59   \cr
Teacup AGN & SDSS J143029.88+133912.0  & 2 & 0.0852 &1.58   \cr
UGC 7342 &  SDSS J121819.30+291513.0 & 2 & 0.0477 & 0.98  \cr
UGC 11185 & SDSS J181611.61+423937.3 & 2 & 0.0412 & 0.80  \cr
\enddata
\end{deluxetable}

\subsection{WISE data}

Our estimates of the bolometric luminosity of the AGN, and hence the role of
obscuration on the ionizing radiation, can be updated from \cite{mnras2012} using
the more sensitive and higher-resolution survey data from WISE \citep{WISE}
in place of IRAS or {\it Akari} for the mid-infrared region including 22 $\mu$.
WISE data show that much of the mid- and far-IR flux attributed
initially to UGC 7342 at $z=0.047$ comes instead from a background starburst system
at $z=0.069$ (as identified in Paper I). The AGN luminosity we derive for UGC 7342 is correspondingly lower,
and the energy deficit for ionizing its gas filaments becomes even stronger.

The WISE catalog magnitudes are listed in Table \ref{tbl-wise}. For further use, we converted them
into fluxes using the zero points from \cite{WISEsupplement}
for ``compromise" spectral slopes.

The WISE data are
especially important since much of the reradiation from circumnuclear dust 
occurs in the mid-IR, which was poorly covered by previous surveys. The infrared output of these galaxies, and thus the potential fraction of obscured and
reradiated AGN radiation, can be quantified in several ways. We have derived total
fluxes and luminosities by integrating power-law spectra connecting the data points and
extrapolating to the boundaries of common bands, and by fitting the data to various
template
spectra energy distributions (SEDs) derived for AGN. Table \ref{tbl-wise} includes mid-IR
luminosities. These are given as the log of the value in erg s$^{-1}$ calculated by power-law interpolation 
between the WISE bands, 
with a redward extrapolation to span from 3.4-42
$\mu$m for comparison with the IRAS-based FIR luminosity as listed in \cite{mnras2012}, which 
includes the range
42-122 $\mu$m \citep{FullmerLonsdale}. SDSS 1510+07 is undetected in the longest WISE band; the
listed luminosity assumes the true flux is at this limit, while setting the flux to zero gives a 
luminosity 0.14 smaller in the log. For comparison, we also list the L(FIR) value (likewise, decimal log 
of the value in erg s$^{-1}$) from \cite{mnras2012}. Many of these objects have mid-IR luminosities
much greater than the FIR values, high enough in Mkn 1498 and the Teacup system to leave open the possibility of an obscured rather
than fading AGN in the absence of other evidence (such as the recombination histories we calculate below). In the others, even including the MIR range often most sensitive to AGN heating of surrounding grains, 
an order-of-magnitude energy mismatch is seen between the AGN itself and the requirement to ionize distant clouds. For UGC 7342 and SDSS 2201+11, the shortfall exceeds a factor 100.
These luminosity estimates are considered in comparison to ionization-based values in
\S 4.3.

\begin{deluxetable}{lccccccc}
\tablecaption{WISE magnitudes and IR Luminosities\label{tbl-wise}}
\tablewidth{0pt}
\tablehead{
\colhead{Galaxy} & \colhead{W1 (3.4$\mu$)}& \colhead{W2 (4.6$\mu$)}  & \colhead{W3 (12$\mu$)} & \colhead{W4 (22$\mu$)}
& \colhead{Log L(MIR)} & \colhead{Log L(FIR)}}
\startdata
IC 2497 & $11.46 \pm 0.02$ &  $11.15 \pm 0.02$ & $7.31 \pm 0.02$ & $4.54 \pm 0.03$ & 44.28 & 44.77 \\
Mkn 1498 & $10.26 \pm 0.02$ &	$9.24 \pm 0.02$ & $6.31 \pm 0.02$ &  $3.72 \pm 0.02$	& 45.41 & $<43.70$ \cr
NGC 5252 & $9.19 \pm 0.02$ & $8.38  \pm 0.02$ & $6.35 \pm 0.01$ & $4.47 \pm 0.02$  & 44.28 & 43.60  \cr
NGC 5972 & $11.02 \pm 0.02$ &$10.49 \pm 0.02$ & $7.05 \pm 0.02$ &	$4.61 \pm 0.02$ & 44.23 & $<43.74$ \cr
SDSS 1510+07 & $12.93 \pm 0.04$ & $12.81 \pm 0.03$ & $11.46 \pm 0.12$ & $>9.00$  & 43.15 & $<44.60 $ \cr
SDSS 2201+11 & $11.10 \pm 0.02$	 & $11.08 \pm 0.02$ & $8.60 \pm 0.02$ & $6.77 \pm 0.07$	& 43.57& $<43.78$  \cr
Teacup AGN & $11.67 \pm 0.02$ & $10.50 \pm 0.02$ & $7.14 \pm 0.01$ & $4.16 \pm 0.02$	  & 45.42 & $<43.36$ \cr
UGC 7342 &	$12.70 \pm 0.02$ & $12.61 \pm 0.03$ & $9.90 \pm 0.05$  &  $7.17 \pm 0.10$	& 43.70 & $<44.04$ \cr
UGC 11185 &	$11.22 \pm 0.02$ & $10.61 \pm 0.02$ & $7.85 \pm 0.02$ & $4.88 \pm 0.02$	& 44.53 & $<44.25$ \cr
\enddata
\tablecomments{WISE data are magnitudes in filters W1--W4. Luminosities are computed for the ranges 3.4-42 $\mu$m for FIR and 42-122 $\mu$m for FIR, expressed as the decimal log of the values in erg s$^{-1}$.}
\end{deluxetable}

\subsection{Spectroscopic mapping}

We investigate the kinematics and ionization structure near the nuclei of several of
these galaxies using integral-field optical spectroscopy from the Gemini Multiple-Object
Spectrometer (GMOS) system \citep{Davies1997}
at the 8m Gillette Gemini-N telescope on Mauna Kea, under programs 
GN-2013B-Q-25 and GN-2014A-Q-25.
The GMOS integral-field unit (IFU) samples the sky in 0.2\arcsec\ apertures covering
a region roughly $3.5 \times 5$\arcsec\ in the single-slit mode we used \citep{AllingtonSmith}.
The B600
grating covered a wavelength range usually encompassing from He II $\lambda 4686$
to the [S II] $\lambda \lambda 6717, 6731$ lines, adjusted for each object's
redshift, at 0.468 \AA\  per pixel. Three stepped grating settings, usually at 200-\AA \ intervals, were 
used to fill gaps between the three GMOS CCDs, with two 20-minute exposures at
each step to identify cosmic-ray events. Reduction of these data followed the
procedures described by \cite{Davies2015},
including correction for internally-scattered light and use of LACOSMIC \citep{lacosmic}
to reject pixels contaminated
by cosmic rays. The RMS scatter of measured comparison-line wavelengths across the
whole array was 0.08 \AA\ ,  5 km s$^{-1}$ for [O III], so the accuracy of this calibration is much finer than the amplitudes of velocity structures we trace.
Data from each observation were combined into a data cube evenly
sampled on the sky, taking into account atmospheric dispersion. Data cubes from all three grating positions were combined at the
end to provide an additional level of rejection for cosmic rays and other 
sources of bad pixels. Flux calibration used a single observation of HZ 44; as the
observations occurred on dates from 13 August 2013 to 27 August 2014, this
flux calibration will be only approximate. We do have external checks on 
line ratios based on previous long-slit spectroscopy (\citealt{mnras2012}, \citealt{paper1}).
Both UGC 11185 data runs were truncated by deteriorating conditions; we have only 
a single grating tilt for each IFU position, with $2 \times 1200$-second exposures, and these are
affected significantly by cosmetic issues, so coverage of some spectral lines is incomplete
and their flux scales are unreliable.

The delivered image quality ranged from 0.4-0.8\arcsec\ FWHM during these
observations, as estimated from red-light ($r$-band) acquisition images obtained immediately before the spectral data. Most of these observations had
FWHM $<0.5$\arcsec. As listed in Table \ref{tbl-ifudata}, we have observations centered on
the nuclei of Mkn 1498 and NGC 5972, and sets of 2 overlapping offset fields
for UGC 7342 and 11185; the table includes the requested offsets from the galaxy nucleus to the observed IFU center point. For the Teacup AGN, three offset fields cover
virtually all the emission-line regions seen in the HST ACS images \citep{paper1}.
Similar data for this galaxy taken with the ESO VIMOS system have been
shown by \cite{Harrison}; they describe results only for the [O III] emission lines, and the
spatial scale of 0.67\arcsec\ per pixel limits their spatial resolution compared to
these GMOS data.

\begin{deluxetable}{lcccl}
\tablecaption{Gemini GMOS IFU spectra \label{tbl-ifudata}}
\tablewidth{0pt}
\tablehead{
\colhead{Location} & Offsets (\arcsec)& \colhead{UT dates} & \colhead{FWHM (arcsec)}  & \colhead{Observers}}
\startdata
UGC 11185 pos 1 & 0.2 W 3.4 S  & 13, 27 August 2013 &   0.48 &    M. Hoenig\\
UGC 11185 pos 2 & 2.5 E 1.2 S  & 13 August 2014    &   0.41 &     A.-N. Chene\\
UGC 7342  pos 1 & 1.7 N & 26 June 2014     &    0.79 &      J. Chavez\\
UGC 7342  pos 2 & 1.7  S & 24 June 2014   &      0.76 &     M. Pohlen\\
Mkn 1498        & 0 & 26 July 2014       &  0.40 &     L. Fuhrman\\
NGC 5972       &  0.3 E & 26 June 2014    &     0.47 &     J. Chavez\\
Teacup ENE    &   4.02 E 2.30 N & 27 July 2014    &     0.56 &    L. Fuhrman\\
Teacup E       &  2.78 E 0.04 N & 18 July 2014      &   0.44  &     M. Pohlen\\
Teacup SW    &   1.5 W 0.63 S &  30 June 2014    &     0.45 &    J. Ball\\
\enddata
\end{deluxetable}

For straightforward comparison of emission-line properties over wide
ranges in signal-to-noise ratio (SNR), we fit Gaussian profiles to the
lines, constraining H$\alpha$ and the adjacent [N II] lines to have the
same full width at half-maximum (FWHM), with the same constraint on the 
[S II] doublet. In some regions, each line is double or triple, so we fit
blended sets of Gaussians; we used the IRAF task {\tt splot} interactively
on the data cubes sampled in 0.2\arcsec\ spatial increments. Line properties of
interest (flux, wavelength, FWHM) were gridded into spatial maps
for further analysis. In most cases, two-component fits were constrained to have the 
same FWHM, for consistency across a wide range in SNR; for the narrower profiles dominated by
the instrumental resolution, this is a good fit. Elsewhere, there may be systematics
introduced, but to first order the flux in each component is preserved.

\section{Ionization Geometry}

Some of these cloud systems show plausible ionization cones, briefly mentioned in Paper 1.
In Fig. \ref{fig-cones} we show line-ratio images of [O III]/H$\alpha$, which often map these structures more clearly than [O III] intensity alone. Among our galaxies, some show
cone-like structures in line intensity; we see a few where the line ratio maps ionization in
a clearer structure. Most notably, in the Teacup AGN there is a reasonably well-defined triangular
region of higher ionization (where the ``hole" lies within it). In others, such as UGC 7342
and UGC 11185 \citep{Hainline}, the entire
emission region could plausibly show a biconical structure without any internal
ionization differences. In these two systems, radial changes in the line ratio are greater than
azimuthal variations at fixed projected radius.
 
 Broadly, all these emission complexes show a twofold symmetry, as sketched in
 \cite{mnras2012} from early data. When their outer parts resemble ionization
 cones, we often see much broader regions near the core (often completely encompassing the
 AGN), whose ionization may have a distinct mechanism. 
  
 The emission region in Mkn 1498 is highly elongated but not biconical. Among the remainder,
 most of the opening angles implied by the HST data are close to our initial estimates, except for
 the Teacup AGN where we now recognize an inner structure traced by the [O III]/H$\alpha$ ratio
 for which each half has a full angle of 37$^\circ$. In SDSS 2201+11, we derive 25$^\circ$,
 although this may be affected by obscuration within the host galaxy on one side of each
 cloud. The rest are broader, with NGC 5252 at $64^\circ$, UGC 7342 at $75^\circ$, 
 and NGC 5972 at $96^\circ$ to encompass the outer structures. The broadest are seen in
 UGC 11185 ($116^\circ$) and SDSS 1510+07 ($126^\circ$). Projection effects in
 a conical distribution make this angle appear larger than its three-dimensional value.
 
 \begin{figure*} 
\includegraphics[width=140.mm,angle=0]{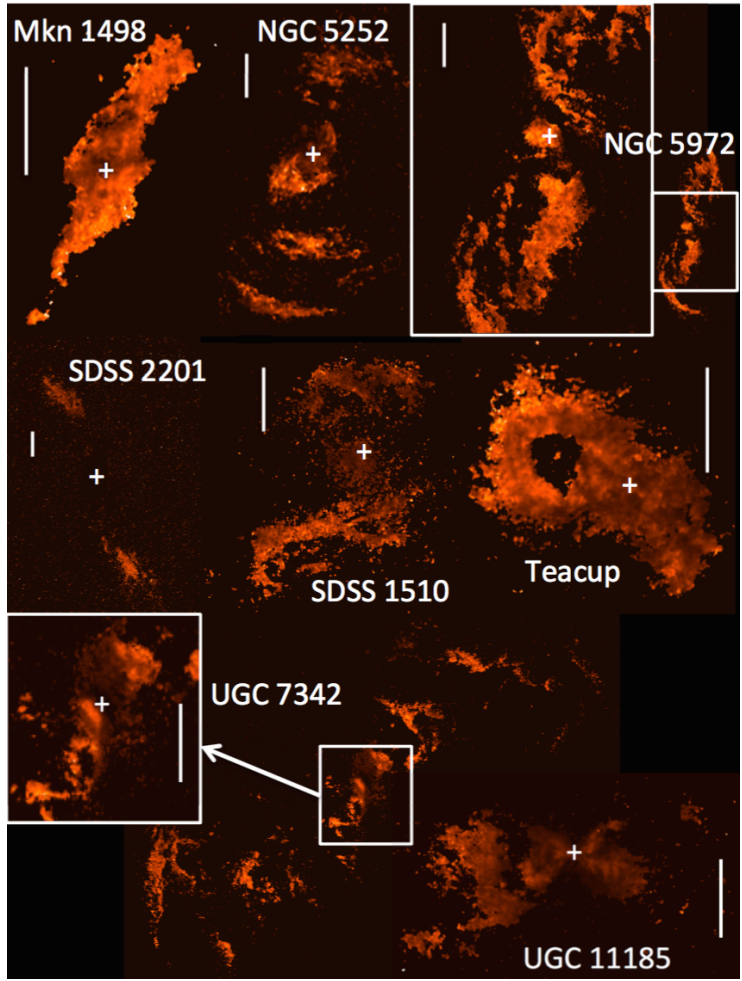} 
\caption{[O III]/H$\alpha$ line-ratio images, using a ``blackbody" color palette to increase the
visual dynamic range and to show features dimmer than easily appear with a pure
grayscale. The scale is the same for all, running from zero to a line ratio of 3.6 shown as white.
The images were masked at 2$\sigma$ in H$\alpha$ intensity (the weaker line in
most areas), then a $3 \times 3$-pixel median filter was used to suppress noise since
the line ratio is more consistent across pixels than the intensities.
North is at the top and east to the left; the angular scale varies as shown by 5\arcsec\
scale bars. For the large cloud systems of
NGC 5972 and UGC 7342, insets show where the enlarged nuclear regions are located.
The nuclear region of SDSS 2201 is affected by reddening from a dust lane, so no reliable
line ratios are available in the inner parts. The location of each AGN is shown by a plus sign.
} 
\label{fig-cones} 
\end{figure*}

Side-to-side asymmetries in ionization (as in BPT plots, or in reconstructed
ionization history) could in principle 
record front-back light-travel-time differences when seen at a sharp angle to the 
axis of ionization cones or radially elongated clouds. We don't actually see any such
differences; our selection for large transverse extent favors cloud systems viewed near the
plane of the sky.
		
In SDSS 2201+11, a dust lane is continuous with the emission-line material into the northern
cloud, showing that the northern cloud must be on the front side of the system 
in order to give significant dust attenuation.  

Near the nuclei, and outside the ionization cones, several of these galaxies show
gas with low ionization as seen in [O III]/H$\alpha$. One goal of the GMOS IFU spectra
was to measure additional line ratios in these regions to probe their ionization mechanism.
We will explore this issue elsewhere, in conjunction with recent HST STIS spectra isolating key 
regions near selected nuclei.

Overall, the occurrence of ionization cones pointing to the nucleus suggests that the
ionized gas has distributions which are mostly radial, rather than, for example, ringlike.
This means that projected distance from the nucleus can serve as a useful proxy for 
physical separation from the AGN in these clouds.

\section{AGN Luminosity History}

\subsection{Method: recombination balance}

We explore here the use of recombination-balance arguments to provide lower limits to the 
ionizing flux reaching various parts of the AGN emission regions. Except in
regions where we see superposition of multiple emission region along the line of sight,
or similarly, elongated filaments seen end-on, the measured luminosity in
a recombination line gives a lower limit to the ionizing luminosity seen by that
volume of material. The brightest pixels at each projected radius give the
most stringent limits. Even in these locations, the derived luminosity will be an underestimate
because of the combined effects of optical depth and covering fraction within the
region spanned by each pixel on the object; both quantities are likely to have
complex and patchy distributions. Suitably located gas must be present to apply this approach, 
of course; a strongly flaring AGN would look the same as one without suitable neutral gas
around the host system.

Despite these clear limitations, this approach could
yield useful information on for example, the history of AGN luminosity, if the 
amount of underestimation is comparable for the highest pixels at various radii (if not necessarily between objects). Regions farther from the AGN are physically unlikely to intercept progressively more radiation, since the density in emission regions general falls with distance from the core, and the changes we will infer span orders of magnitude in the opposite sense to be expected from this. 
Two immediate checks on this approach are available - the
photoionization modeling for spectra of various region around the Teacup AGN \citep{Gagne}, 
and narrowband [O III] structures of a set of Palomar-Green QSOs.

We employ the ionization balance as follows. We use the highest surface brightness observed in H$\alpha$ as a function
of projected distance from the AGN to estimate the required (isotropic)
emission rate of ionizing photons $Q_{ion}$ needed to power the observed emission,
as noted above making the broad assumption that the densest gas clumps are similar at 
various projected distances 
from the AGN within each system.  In practice, we use the required
photon rate between the H and He ionization edges, since the
photoionization cross-section of helium is large enough to dominate
the absorption immediately above 54.6 eV. As expected, the HST images reveal
numerous small regions of high surface brightness, so that the
lower limit we derive in this way is often much higher than
estimated from ground-based spectra \citep{mnras2012}. Our 
calculation incorporates the fraction of recombinations leading to
an H$\alpha$ photon  (\citealt{SpitzerGreenstein}; \citealt{Nussbaumer}). 
In ``Case B", where the gas is optically thick to Lyman $\alpha$,
the fraction of recombinations leading to an H$\alpha$ photon
ranges roughly from 0.25-0.34 depending on the trapping properties
of Lyman $\alpha$. The other limit, case A where the nebula is optically
thin in Lyman $\alpha$, has these fractions lower by a factor 1.6 
(these values are for $T_e = 10^4$ K, with recombination coefficients 
as collected by \cite{AGNAGN2006}). If we denote
the fraction of recombinations leading to an H$\alpha$ photon by $f$,
$Q_{ion}$ can be derived conveniently from the area-normalized
count rate $N$. This value represents the arrival rate (Hz cm$^{-2}$) of line
photons in the imaging data, corrected for the quantum efficiency $\eta$
($ \approx 0.39$ of the  
HST+ACS+filter system at H$\alpha$; for NGC 5972, where H$\alpha$ 
was observed with WFC3 and F673N filter, the value is 0.22; for NGC 5252
with the WFPC2 ramp filter, 0.119). The current HST pipeline calibration corrects for the gain
(yielding results in electrons per second $N$ , which is to say detected photons) for ACS and WFC3 data, but not for WFPC2 data (so the gain factor must be applied to these images);
the incident flux of H$\alpha$ photons is then $N A/ \eta$ for unobscured telescope area $A$,
 so the number emitted
per second in the emitter's frame is $ 4 \pi D^2 (1+z) N A / \eta$ for source distance
$D$, now including the time dilation $(1+z)$ in photon arrival rate.
Modulo projection effects which are
poorly known, material in each pixel of observed solid angle $a$ 
would intercept a fraction of the emitted
radiation of order $a / 4 \pi R^2$ where $R$ is the projected 
distance from the nucleus to the area of interest; if $R$ is in 
pixels, $a=1$ for unsmoothed data. 

Collecting these
components, the count rate $N$ per unit area 
in a single pixel implies
$$ Q_{ion} =  (1+z) (N / \eta f)  4 \pi D^2 4 \pi R^2 $$ 
for a region projected $R$ pixels from the AGN. For the density and 
Lyman $\alpha$ optical depth ranges that matter here, as noted
above, $f$ ranges from 0.16-0.34. This value of $Q_{ion}$ is a lower
limit, since we then have no constraint on how much radiation escapes 
the system. To be conservative, we discuss values using $f=0.29$, in the middle
of the range of Case B photoionization.  
This was cast in a somewhat different way in \cite{mnras2012}, using 
the angular width of a spectrograph slit about the nucleus at a
given $R$.

The ACS ramp-filter passband has a nearly flat profile near the peak
wavelength, falling below 50\% of peak transmission well outside the
span of the [N II] lines when centered on H$\alpha$ at these redshifts.
We make a correction for the adjacent [N II] emission based on our 
slit spectroscopy (Keel et al. 2012), assuming the line ratio behavior
to be circularly symmetric about the nuclei on each side. We fit
this correction factor as a polynomial in radius for computational convenience.

For optically thin gas, or material with a small covering factor 
(when shadowing by
absorbers closer to the nucleus is not important), there will be an upper envelope
in the H$\alpha$ surface brightness-projected radius diagram,
which will follow a $1/r^2$ form for constant source luminosity
as long as various regions in emission do not have radically different
projection factors. Departures from this behavior can be evidence of long-term 
changes in $Q_{ion}$; neglecting projection factors, one can in 
principle extract the source history from such a diagram.

\subsection{Consistency tests}

We show one set of data for this calculation suitable for an external test in 
Fig. \ref{fig-teacupharad} for the Teacup
AGN, where the surface brightness is in electrons pixel$^{-1}$ s$^{-1}$ and the
radius is in arc seconds. Transforming these into
the physically meaningful units of $Q_{ion}$ and projected 
distance in light-years gives our estimates of the minimum $Q_{ion}$
at each pixel.

\begin{figure*} 
\includegraphics[width=125.mm,angle=270]{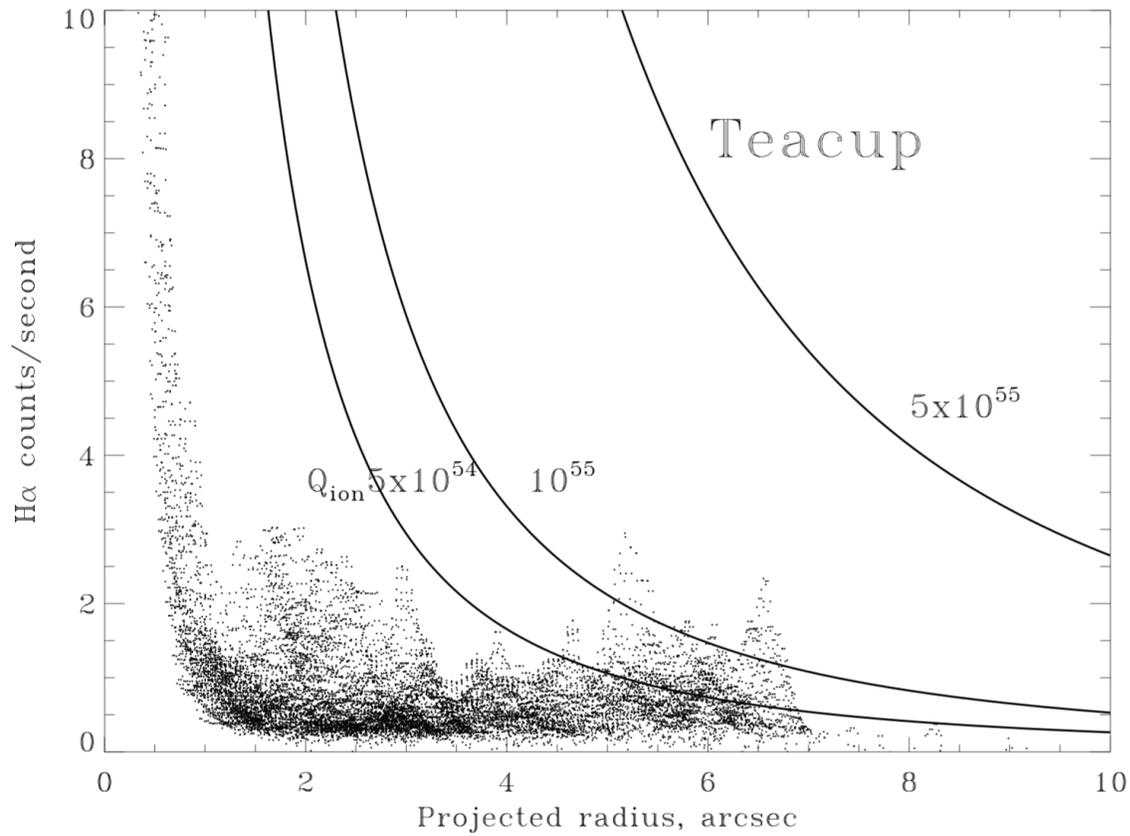} 
\caption{Radial plot of pixel-by-pixel surface brightness (in instrumental units, detected 
photons per second per pixel)
in H$\alpha$ for the Teacup AGN, with superimposed curves of constant $Q_{ion}$ in photons second$^{-1}$. Median filtering over $3 \times 3$ pixels was used to suppress residual cosmic-ray and charge-transfer effects, with the side effect of creating
correlations between some sets of adjacent points.} 
\label{fig-teacupharad} 
\end{figure*} 

This conversion is illustrated in Fig. \ref{fig-teacupqplot}, showing $Q_{ion}$ 
converted into ionizing luminosity for each pixel above a $3 \sigma$
threshold for the Teacup AGN. The upper
envelope is the most significant feature, since we argue that the
densest regions are likely to be comparable at various radii. For this object,
we can compare our results to the detailed photoionization
modeling using the CLOUDY 90 code \citep{CLOUDY90} carried out by \cite{Gagne}
using ground-based spectra spanning its emission regions. Those results parallel the
upper envelope of our reconstructed points, systematically $\approx 25$\% higher
depending on how the envelope is defined,
which supports both our results and the basic assumption of our analysis, that 
the most optically-thick regions are comparable at all radii in the clouds (so we tentatively assume
this to be the case for the other galaxies in our sample). In particular, our analysis reproduces the
same change in luminosity, declining by by a factor $10^{1.03}$ over a time $\Delta T = 26500$ years
between the peaks in reconstructed luminosity. It is also reassuring that the peaks in reconstructed
luminosity approach the spectroscopic results closely but do not exceed them at any point.

\begin{figure*} 
\includegraphics[width=125.mm,angle=270]{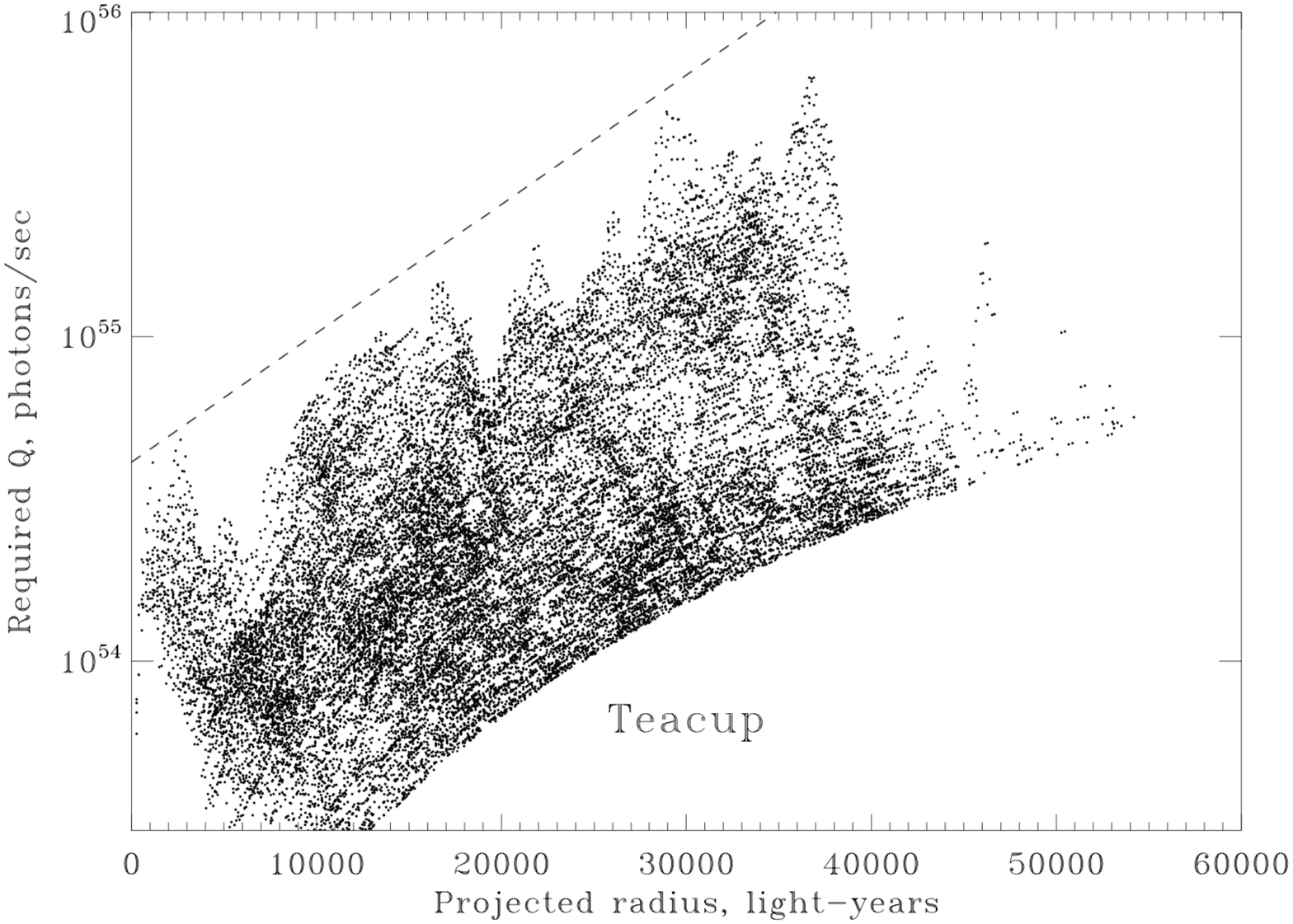} 
\caption{Radial plot of pixel-by-pixel minimum required ionizing luminosity for the Teacup AGN,
derived from the data in Fig. \ref{fig-teacupharad}. 
Projected distance from the AGN is expressed in light-years to make the timescales apparent.
The dashed line at the top represents the ionization history inferred independently by \cite{Gagne}
from detailed modeling of ground-based spectra. The envelope at the bottom represents
a 2-$\sigma$ cut to suppress the effects of pixel noise at large radii} 
\label{fig-teacupqplot} 
\end{figure*} 

As a second test of our approach, we use HST data on low-redshift QSOs for 
their high spatial resolution, since numerical tests on our images show that the 
derived peak
ionizing flux drops rapidly with lower image quality. (If this behavior proves to be consistent
enough, an analogous study could be done on the emerging body of integral-field
spectroscopic mapping of AGN with EELRs, such as \citealt{Liu} or \citealt{Husemann}). 
We compare
with the narrowband [O III] data shown for 7 Palomar-Green QSOs by \cite{Bennert2002},
which use the WFPC2 linear-ramp filters and continuum subtraction incorporating stellar
PSFs. The scaling of their implied ionizing luminosities is only approximate; we used a mean
value for 
[O III]/H$\alpha =4$, lacking independent narrow-line ratios at the core. That program
was targeted at NLR sizes, with no selection for extent of EELRs, and only three of the PG QSOs 
have detected emission beyond the 10-kpc radial bound used to select our fading-AGN sample.
We masked regions around obvious companion galaxies, but not a bright emission region
within 1\arcsec\ of PG 0157+001 since its nature is ambiguous.
The implied histories are shown in Fig. \ref{fig-pghistory}. These show brightening, fading,
and near-constant episodes. PG 0157+001 and PG 1012+008 show fading by nearly
an order of magnitude before brightening close to the observed epoch, while PG 0953+414
has brightened by nearly this much. Brightening episodes will be harder to detect at large
radii as the signal drops more strongly near the detection threshold. This test broadly
confirms that our sample of fading candidates does indeed show distinct behavior 
using our luminosity reconstruction technique from recombination balance. In particular,
{\it none of the PG QSOs show the rapid drops in the last 20,000 years that are
ubiquitous in our fading-AGN sample}.

\begin{figure*} 
\includegraphics[width=125.mm,angle=90]{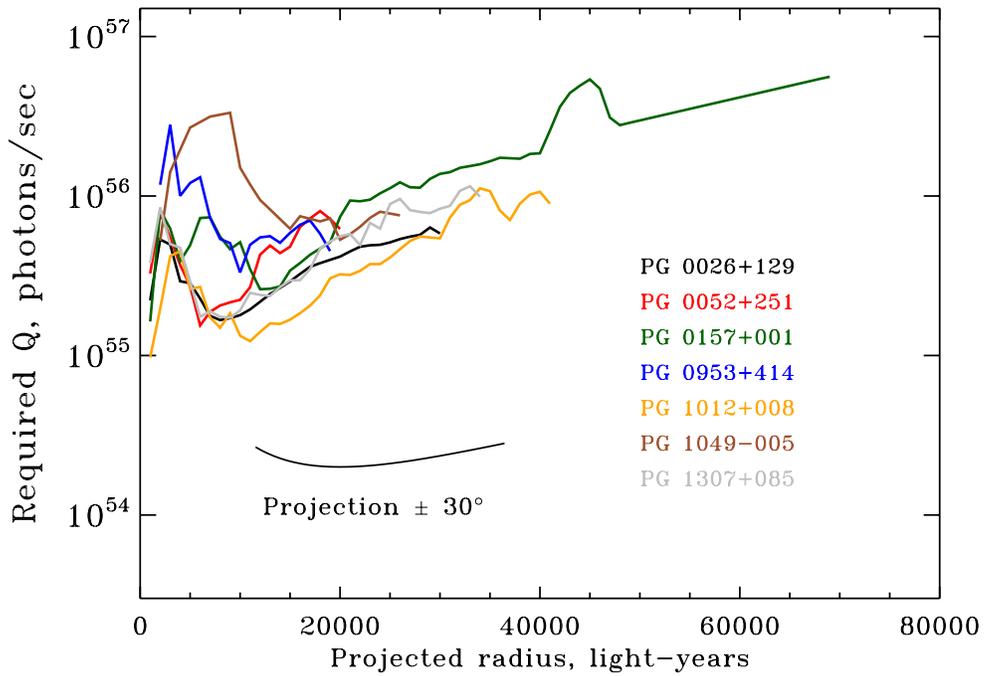} 
\caption{Inferred ionization history of of Palomar-Green QSOs from \cite{Bennert2002},
calculated as in Fig. \ref{fig-historykyrsmooth}, showing peak values in 1000-light-year
bins with radially adaptive smoothing.
 The projection curve at the bottom shows the effect of changing the vector 
direction from the AGN to the gas by
$\pm 30^\circ$ from the plane of the sky, for a measured value at the bottom of this curve.} 
\label{fig-pghistory} 
\end{figure*} 

\subsection{Luminosity histories of fading AGN}

The two external tests outlined above suggest that, indeed, we can reconstruct at least qualitative histories for AGN luminosity using pixel-by-pixel recombination balance. 
Fig. \ref{fig-history2} shows these pixel-by-pixel results for the other 7 objects in our
sample plus IC 2497 and Hanny's Voorwerp, where similar data were described by 
\cite{HSTHV}. Fig. \ref{fig-historykyr} compares the upper envelopes for
all objects, connecting the brightest pixels in 1000-ly bins.

\begin{figure*} 
\includegraphics[width=145.mm,angle=0]{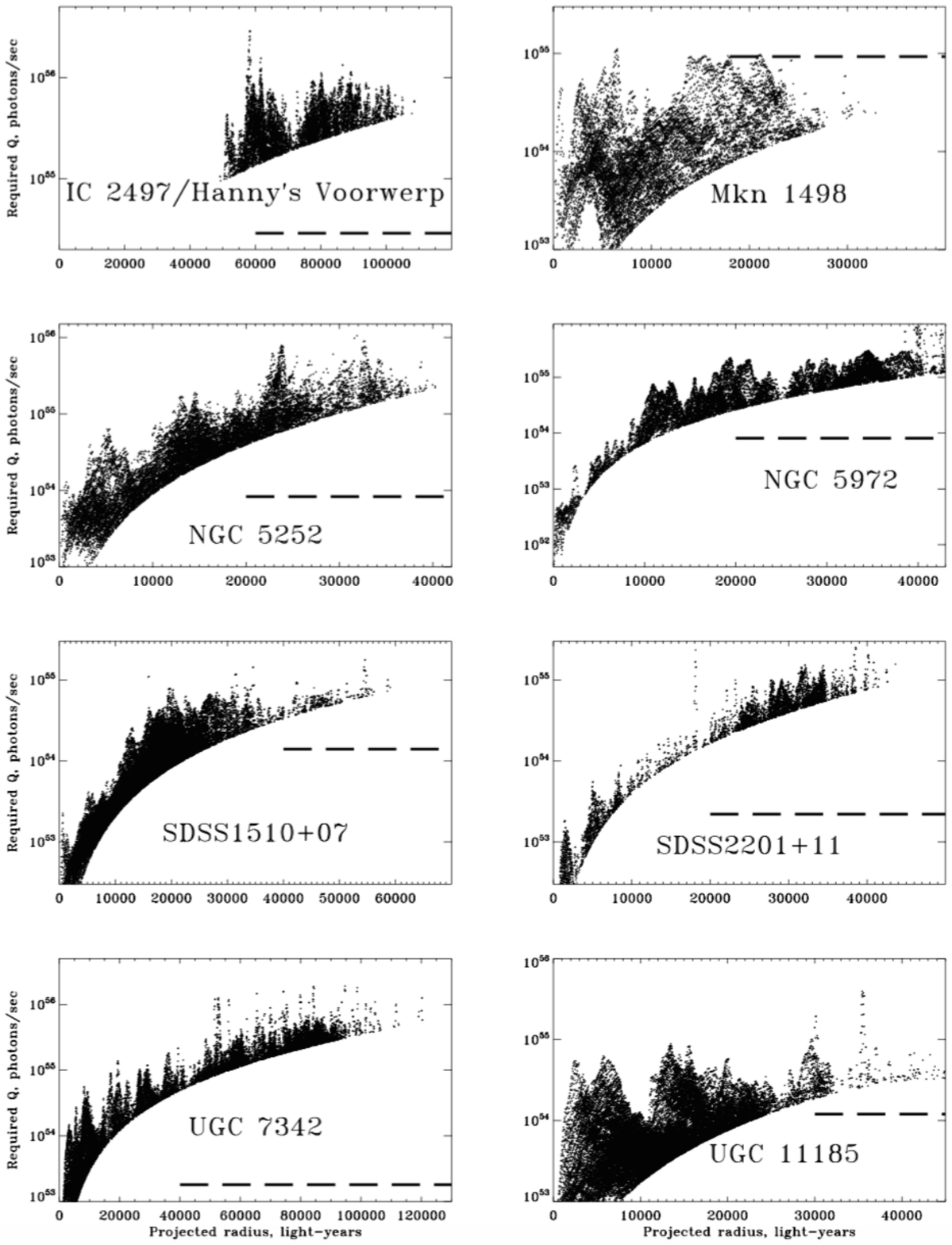} 
\caption{Inferred ionization history of extended AGN-ionized clouds reconstructed pixel by pixel from the surface brightness in H$\alpha$ for each of our sample galaxies, 
as in Fig. \ref{fig-teacupqplot}. The upper envelope is likely to scale with the ionizing luminosity, falling short by a factor related to the optical depth of the material in the brightest pixels (which we assume to be comparable at all radii). Horizontal dashed lines show the ionizing-photon rate inferred from the MIR+FIR luminosity
in the upper-limit case of heavily-obscured AGN; the levels shown for SDSS 1510+07 
and SDSS 2201+11 are themselves 
based on IR upper limits. Values above this cannot be accounted for in a simple case of an 
isotropic AGN heavily obscured along our line of sight.} 
\label{fig-history2} 
\end{figure*} 

Recombination timescales for hydrogen approach
$10^4$ years in the outer parts of these clouds, so only changes on longer terms can reflect variations in the AGN output; along with line-of-sight depth of the cloud systems, this means that the temporal resolution degrades with projected radius. We illustrate this in Fig. \ref{fig-historykyrsmooth}, which smooths the
derived luminosities by a range 0.14 times the radius, 
a value which matches the best-constrained timescale in the outer regions of Hanny's Voorwerp and roughly accounts for a density decrease with radius matching the limits from \cite{cjl2009}. 
The points at $\Delta T = 0$ in both versions of this figure are from the H$\alpha$
fluxes at the nuclei (operationally within a projected radius of 1\arcsec\ from slit spectroscopy), 
in the approximation that they represent full covering of the instantaneous
ionizing flux (strictly, a lower limit since we see that ionizing photons do escape the nuclear region). The flux values are taken from our spectroscopy in \cite{paper1}, as given
in Table \ref{tbl-hanuclei}, including only the narrow component of the composite profile in Mkn 1498 
for consistency. As a rough numerical description of these reconstructed histories, we collect in Table \ref{tbl-timescales}
a set of exponential-fit properties to segments of the smoothed version in 
Fig. \ref{fig-historykyrsmooth}, showing not only fading periods
but constant and brightening episodes.

\begin{figure*} 
\includegraphics[width=125.mm,angle=90]{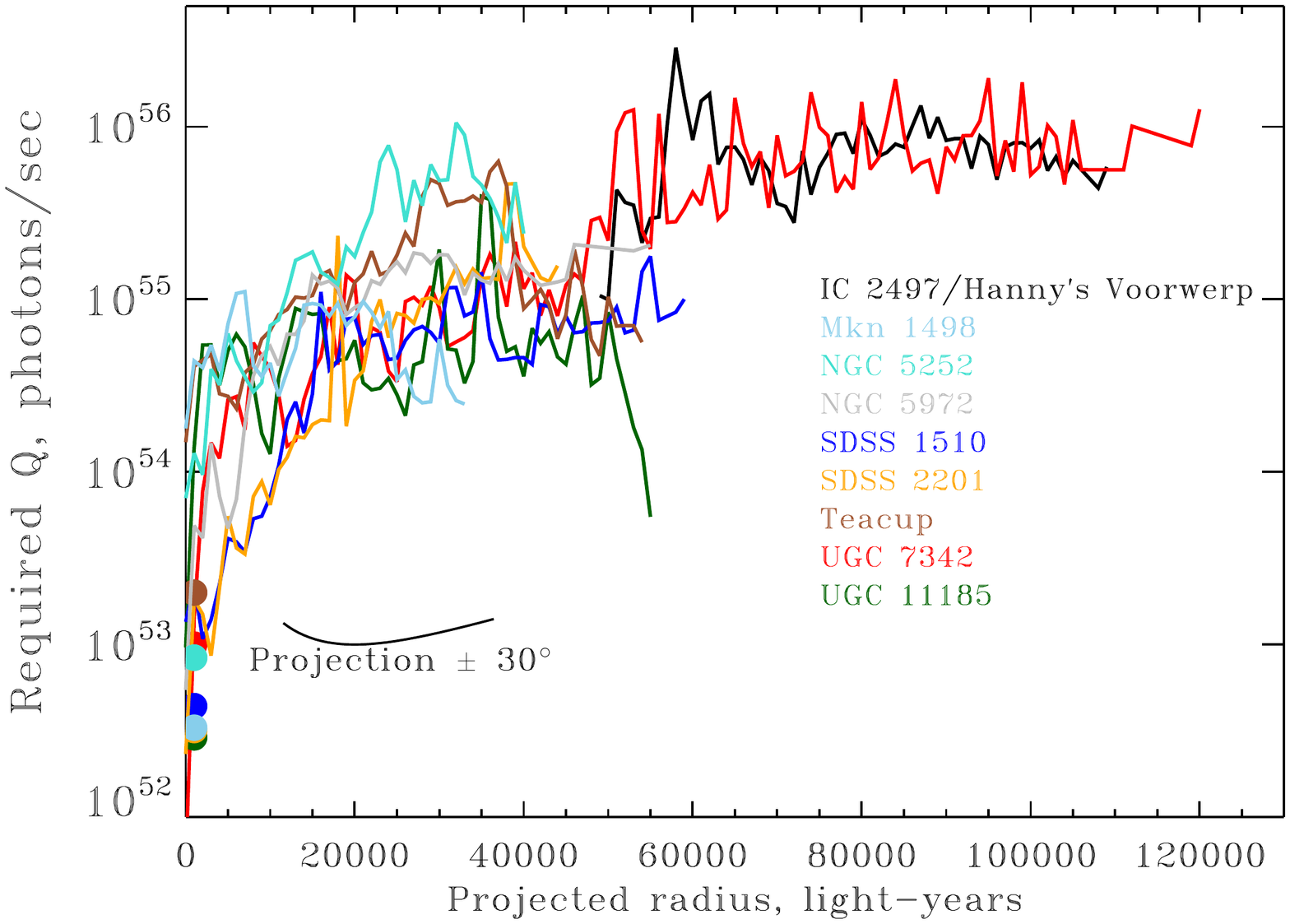} 
\caption{Inferred ionization history of extended AGN-ionized clouds as in Fig. \ref{fig-teacupqplot}, with lines
showing the peak values in every 1000-light-year bin for each object.  The projection curve at the bottom shows the effect of changing the vector 
direction from the AGN to the gas by
$\pm 30^\circ$ from the plane of the sky. The large points near zero radius show the derived
observed-epoch AGN luminosity from H$\alpha$ at the nuclei, except for IC 2497 which falls well off the scale at $1.1 \times 10^{50}$. } 
\label{fig-historykyr} 
\end{figure*} 

\begin{figure*} 
\includegraphics[width=125.mm,angle=90]{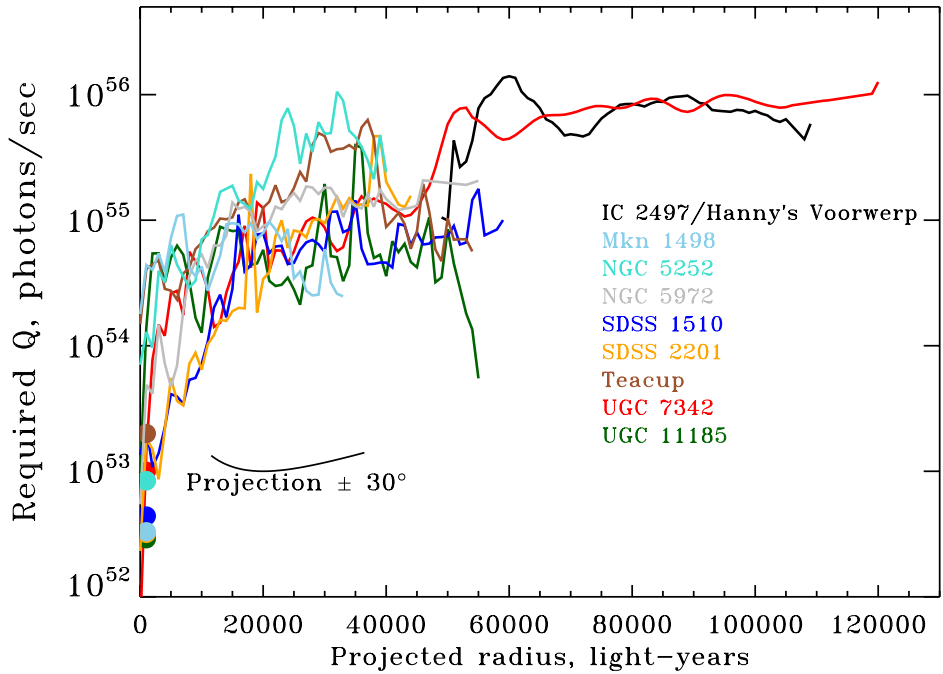} 
\caption{Inferred ionization history of extended AGN-ionized clouds 
showing the peak values in every 1000-light-year bin for each object.  This representation
smooths the derived $Q$ values by radial amounts intended to reflect changes in the
recombination timescale with radius, to better distinguish structure which could be due
to AGN variation rather than gas structure.
The projection curve at the bottom shows the effect on a point at the bottom of the curve,
of changing the vector 
direction from the AGN to the gas by
$\pm 30^\circ$ from the plane of the sky. 
The large points near zero radius show the derived
observed-epoch AGN luminosity from H$\alpha$ at the nuclei, except for IC 2497 which falls well off the scale at $1.1 \times 10^{50}$.} 
\label{fig-historykyrsmooth} 
\end{figure*}

\begin{deluxetable}{lcccc}
\tablecaption{Timescale estimates for variability episodes \label{tbl-timescales}}
\tablewidth{0pt}
\tablehead{
\colhead{Galaxy} & \colhead{Data span (yr)} & \colhead{Time range (yr)}  & \colhead{Behavior} & \colhead{$e$-folding timescale, years} }
\startdata
IC 2497 & 65000 & 111000-102000 & brighten 1.0 dex &  4000 \\
 	&		& 102000-60000 & constant &  --- \\
             &             & 60000-46000 & fade 2.0 dex & 3000 \\
Mkn 1498  & 	32000 & 32000-14000 &  brighten 0.6 dex	&	13000 \\
		&	& 14000-2000 & fade 0.5 dex &  10000 \\
		&	& 2000-0 & fade 1.6 dex & 550  \\
NGC 5252  &  40000    & 40000--14000 & constant	&--- \\
		 &                 & 	14000--2000 &  fade 1.2 dex	& 4300 \\
		&	            & 2000-0   &    fade 1.4 dex	&	620  \\
NGC 5972  &  55000 &	55000-15000 & fade 0.3 dex	& 5800 \\
		&         & 	15000-5000 & fade 1.0 dex	& 4300\\
		&         & 	5000-3000 & brighten 0.5 dex	& 1700\\
		&         & 	3000-0 & fade 1.3 dex		& 1000 \\
SDSS 1510  &  60000 &	60000-18000&  constant	& ---\\
		&     & 	18000-2000 & fade 1.8 dex	& 3900 \\
		&     & 	2000-0    & fade 0.6 dex		& 1500 \\
SDSS 2201   & 45000 &	45000-10000 & fade	1.2 dex	&13000	 \\
		&       & 	10000-3000  & fade 0.8 dex	&	3800\\
		&       & 	3000-0    & fade 0.6 dex		&  2200 \\
Teacup   & 55000 	&	55000-36000  & brighten 1.0 dex	 & 8000\\
		&      & 	36000-1000   & fade 1.2 dex & 12000 \\
(Gagne et al.) &  39000 & 39000 - 0 & fade 1.7 dex & 9800 \\
UGC 7342  & 120000 	& 120000-50000  & constant	& --- \\
		&     & 	50000-34000 & fade 1.03 dex	&	6700\\
		&     & 	34000-20000 & fade 0.1 dex &	60000\\
		&     & 	20000-11000  & fade 0.40 dex  &	9700\\
		&     & 	11000-2000 & fade 0.6 dex  &	6500\\
		&     & 	2000-0    & fade $>$1.0 dex 	 & 900\\
		
UGC 11185 &  55000 	& 55000-45000 & brighten 1.0 dex	&	4300\\
		&     & 	45000-4000  & constant		& ---\\
		&     & 	4000-0 & fade 2.2 dex	& 800 \\
\enddata
\end{deluxetable}

\begin{deluxetable}{lccc}
\tablecaption{Nuclear H$\alpha$ Fluxes and Ionizing Rates \label{tbl-hanuclei}}
\tablewidth{0pt}
\tablehead{
\colhead{Galaxy} & \colhead{H$\alpha$ flux, erg cm$^{-2}$ s$^{-1}$}  & \colhead{Q$_{ion}$, photons s$^{-1}$} & \colhead{MIR+FIR Q$_{ion}$, photons s$^{-1}$}}
\startdata
IC 2497       &  $1.3 \times 10^{-15}$ &    $   2.6 \times 10^{50}$ \\ $2.9 \times 10^{54}$
Mkn 1498     &   $5.5 \times 10^{-14}$ &   $ 3.3 \times 10^{52}$ & $9.3 \times 10^{54}$ \\
NGC 5252    &  $4.5 \times 10^{-13}$   &  $ 8.4 \times 10^{52}$ & $8.3 \times 10^{53}$\\
NGC 5972  &   $2.6 \times 10^{-14}$ &   $     8.4 \times 10^{52}$ & $ 8.1 \times 10^{53}$ \\
SDSS 1510  & $5.9 \times 10^{-14}$ &    $4.4 \times 10^{52}$ & $ < 1.4 \times 10^{54}$ \\
SDSS 2201 &  $1.1\times 10^{-13}$ &   $3.2 \times 10^{52}$ & $< 2.2 \times 10^{53}$ \\
Teacup   &        $  7.9 \times 10^{-14}$ &    $ 2.0 \times 10^{53}$ & $9.5 \times 10^{54}$ \\
UGC 7342   & $ 1.3 \times 10^{-13}$ &   $ 1.0 \times 10^{53}$ & $ 1.8 \times 10^{53}$ \\
UGC 11185 &      $  4.9 \times 10^{-14}$ &   $  2.9 \times 10^{52}$ & $1.2 \times 10^{54}$ \\ 
\enddata
\end{deluxetable}

We considered several alternate approaches to calculating ionizing history, including integrating
the line flux in slices across the putative ionization cones. 
This was defeated by the very patchy structure of the ionized features; tracing only the
peak intensity reduces sensitivity to the larger-scale structure.

We can estimate the values of $Q_{ion}$ predicted from the complementary assumption
that the AGN is unseen along the direct line of sight due to dust obscuration. In this case,
we take the total infrared output to be the total AGN luminosity, and a typical fraction of
the bolometric luminosity 0.16 to be in ionizing radiation from 13.6-54 eV (typical of
composite spectral energy distributions, although this is often the least well-determined piece
of the spectrum). For a wide range
of spectral index in this range, the mean ionizing energy is close to 27 eV \citep{Netzer1990}.
This is intended to be conservative with respect to the possible role of AGN observation, making
no correction for host-galaxy IR contribution or the possibility that significant IR reprocessing
occurs far from the AGN and thus may not reflect its luminosity at the same epoch as our direct view.
The resulting $Q_{ion}$ values or limits for most of these AGN (listed in Table \ref{tbl-hanuclei})
fall well below even the
lower limits imposed by recombination balance, adding to the case for luminosity drops.
Finally, the spatially-resolved drops in required ionizing flux seen within each object's clouds
add to the evidence that the energy
mismatch between nuclei and extended clouds really results from variability rather than
unusually hard FUV spectral slopes or localized extinction around the AGN. The shortfalls for
$Q_{ion}$ derived from the infrared are less extreme, indicating modest local
obscuration, but in most cases still very large.

Most of our sample would not have shown enough fading to satisfy our selection
criteria based only on the early parts of the reconstructed histories. The strongest, characteristic
fading took place within the last 20,000 years before the epoch when we observe the nuclei,
 except for IC 2497/Hanny's Voorwerp, where it occurred in a similar span about
 50,000 years before the direct view and remained dim. This distinctive fading
 is evidence (internal to each galaxy) for actual luminosity drops rather than alternate interpretations of the energy-budget
 mismatch in these objects, such as obscuration from dust far from the nucleus (so there
 is no necessary connection between the obscuration and mid-IR emission) or
 spectral energy distributions unusually strong peaked in the ionizing UV.

\section{AGN Outflows and Feedback}

In IC 2497, host galaxy of Hanny's Voorwerp, HST 
observations reveal an expanding loop of low-ionization gas extending
$\approx 500$ pc to one side of the nucleus \citep{HSTHV}. This,
along with evidence for an oppositely-directed radio jet (\citealt{Jozsa},
\citealt{Rampadarath}) and bubble-like structure in the X-ray gas
\citep{Sartori},
invites speculation that these objects are not necessarily undergoing
a near-shutdown of accretion, but may (also) be switching modes of
energy output incidental to the accretion, to become kinematically dominated
(sometimes known as radio-mode). In X-ray binaries, the switch from radiative to kinetic mode is associated with a change in accretion state (and thus a decrease in the luminosity). It is possible that the same is happening also in AGN \citep{Sartori}. There are additional AGN with
well-studied bubbles or loops attributed to outflows, such as recent results on
NGC 3393 by \cite{Maksym2016}. This motivated us to look for similar
loops of emission, and seek kinematic evidence as to whether they represent outflows or something else.

Morphologically similar loops to one side of nuclei, or encircling them, 
are common in this sample (Fig. \ref{fig-nuchalpha}), and could by themselves be
signatures of outflow episodes. However, the GMOS IFU
data give a more complex picture both kinematically and in emission-line ratios.

\begin{figure*} 
\includegraphics[width=155.mm,angle=0]{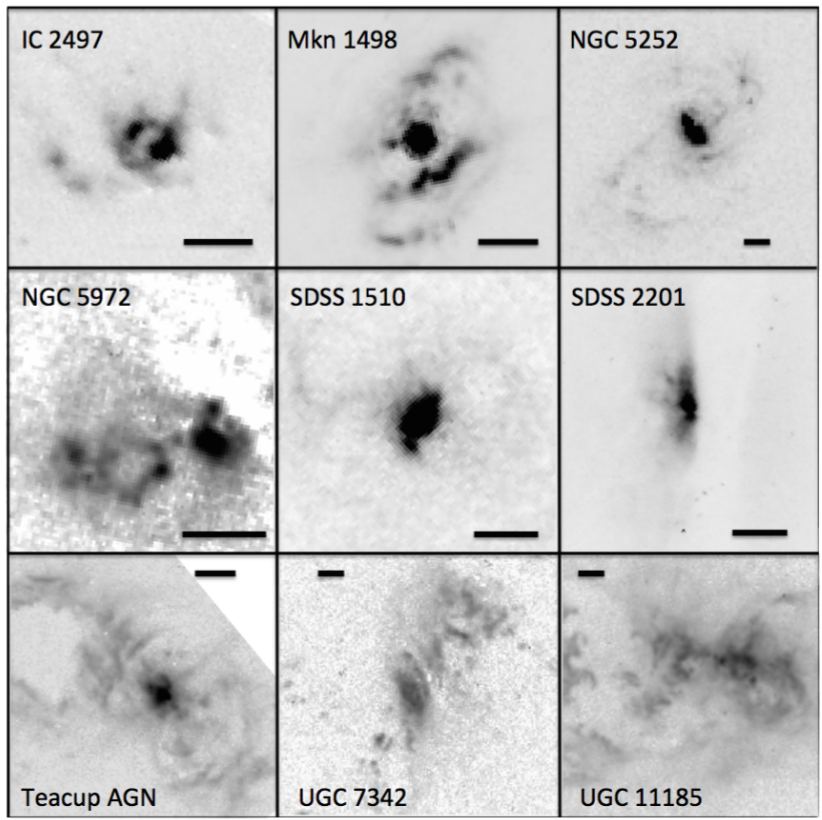} 
\caption{Continuum-subtracted H$\alpha$+[N II] images of nuclear regions, most showing
loops or ringlike emission structures suggesting outflow. In NGC 5972 and SDSS 2201, color 
changes from dust lanes limit the precision of continuum subtraction with the available filters. North is at the top, east to the left; scale bars are 1\arcsec\ long. The three objects in the bottom row
have very bright central emission, so their images are shown with logarithmic intensity
scales; the rest use linear scales.} 
\label{fig-nuchalpha} 
\end{figure*} 

Kinematic signs of outflow do appear in several nuclei, although some of the 
loops are more nearly in rotation than radial motion. We can evaluate this
using the Gemini IFU spectroscopy, as well as the wider-field Fabry-Perot
velocity fields for [O III] obtained with the 6m BTA (Paper I). Spectroscopic
signatures of outflow include multiple-peaked or asymmetric line profiles, and
departures from large-scale rotational velocity fields even for locally narrow and symmetric profiles.
Such features are seen in confined regions of some of our target galaxies.
Fig. \ref{fig-lineprofiles} overlays the [O III] profiles for the Gemini IFU data on
HST narrowband images in [O III], with the line profiles representing averages of the
data  over $0.5 \times 0.5$\arcsec\ regions.

\begin{figure*} 
\includegraphics[width=130.mm,angle=90]{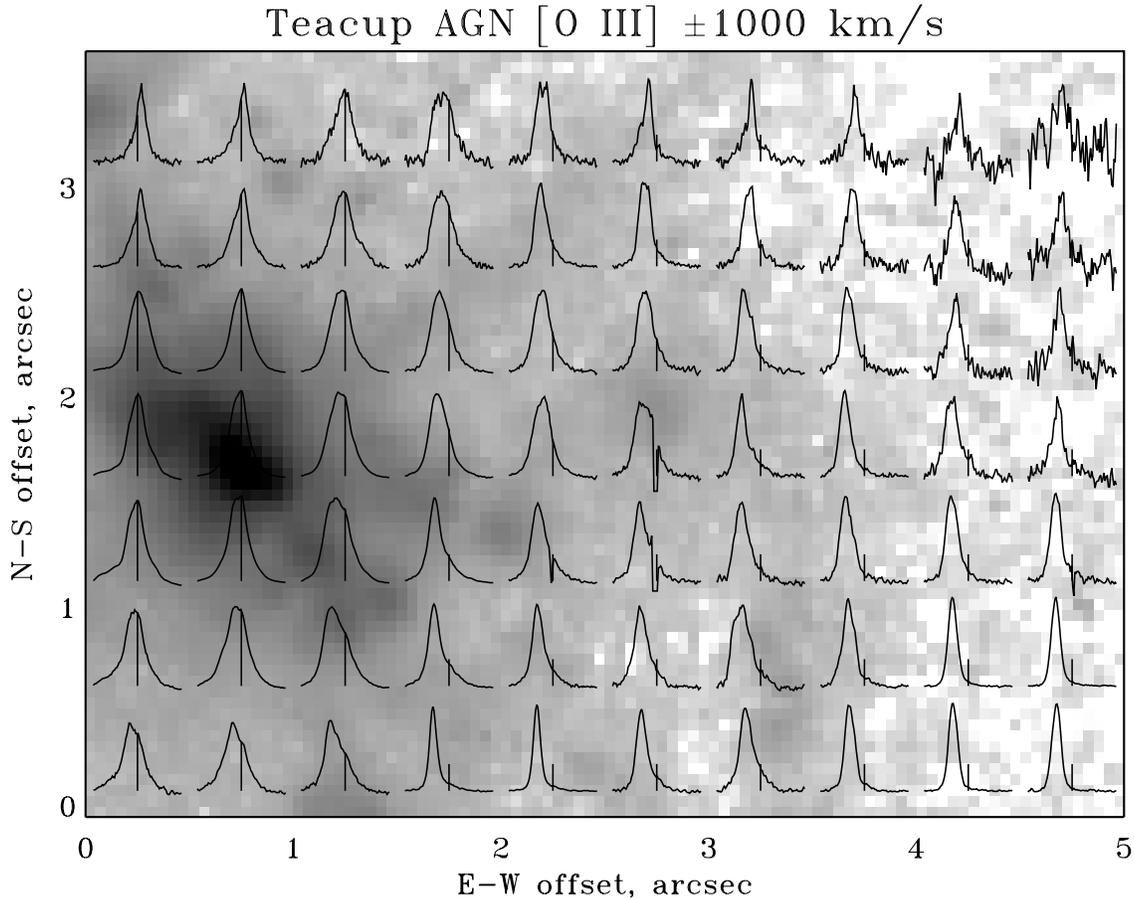} 
\caption{[O III] emission-line profiles from the GMOS IFU spectra overlaid on the HST [O III] images, shown with a common logarithmic intensity scale. Line profiles are from the data averaged over
$0.5 \times 0.5$\arcsec\  regions, and are scaled to have the same peak heights. The velocity range shown is indicated at the top of each panel - $\pm 700$ km s$^{-1}$ for all but the Teacup where we show $\pm 1000$ km s$^{-1}$. The coordinates are in the GMOS IFU field system for ease
of display; this involves 90\arcdeg \ rotations between some of them, and the UGC 7342
regions have south at the top. Alignment 
was set by comparison of smoothed versions of the HST images to monochromatic
slices of the IFU data cubes. For the three Teacup AGN pointings; the first includes the AGN itself, and the next two include the southern and northern parts of the prominent loop to its east. For each line profile, the vertical line marks the systemic velocity
as derived from the nucleus. The Gemini data have excellent sensitivity to diffuse emission,
giving detection of emission even in some areas which are essentially blank in the HST images.
} 
\label{fig-lineprofiles} 
\end{figure*}

\addtocounter{figure}{-1}

\begin{figure*} 
\includegraphics[width=130.mm,angle=90]{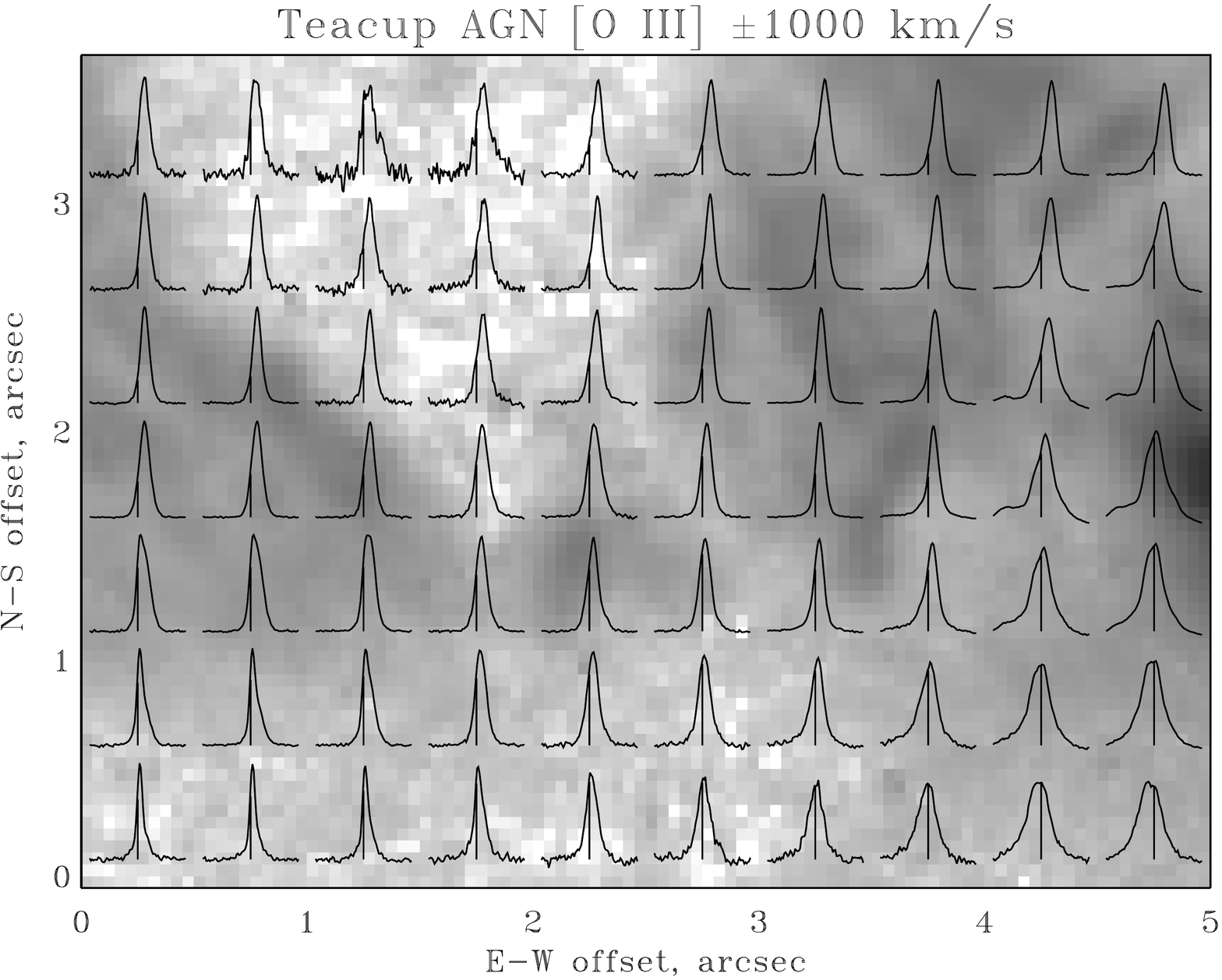} 
\caption{(continued)} 
\end{figure*} 

\addtocounter{figure}{-1}

\begin{figure*} 
\includegraphics[width=130.mm,angle=90]{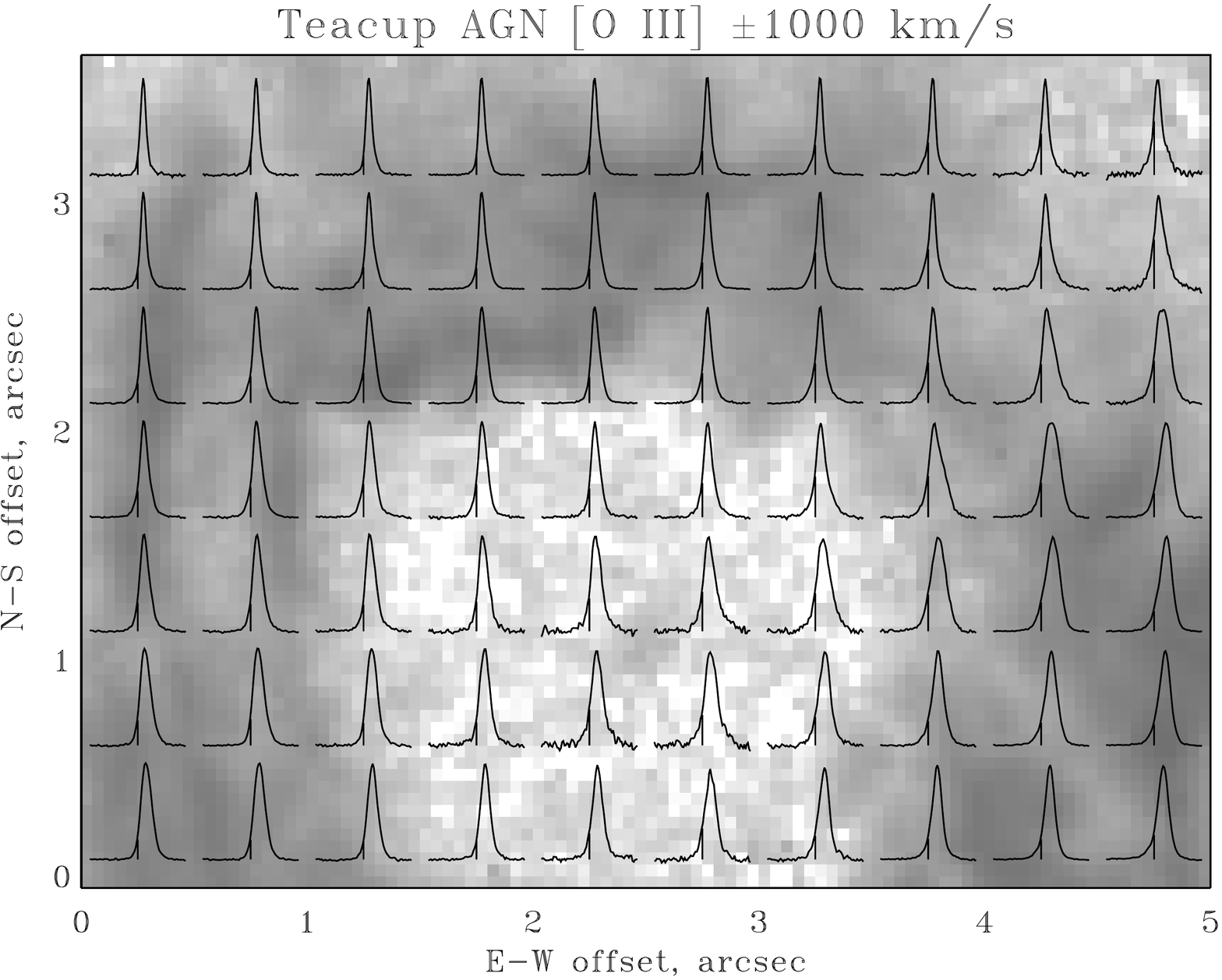} 
\caption{(continued)} 
\end{figure*} 

\addtocounter{figure}{-1}

\begin{figure*} 
\includegraphics[width=130.mm,angle=90]{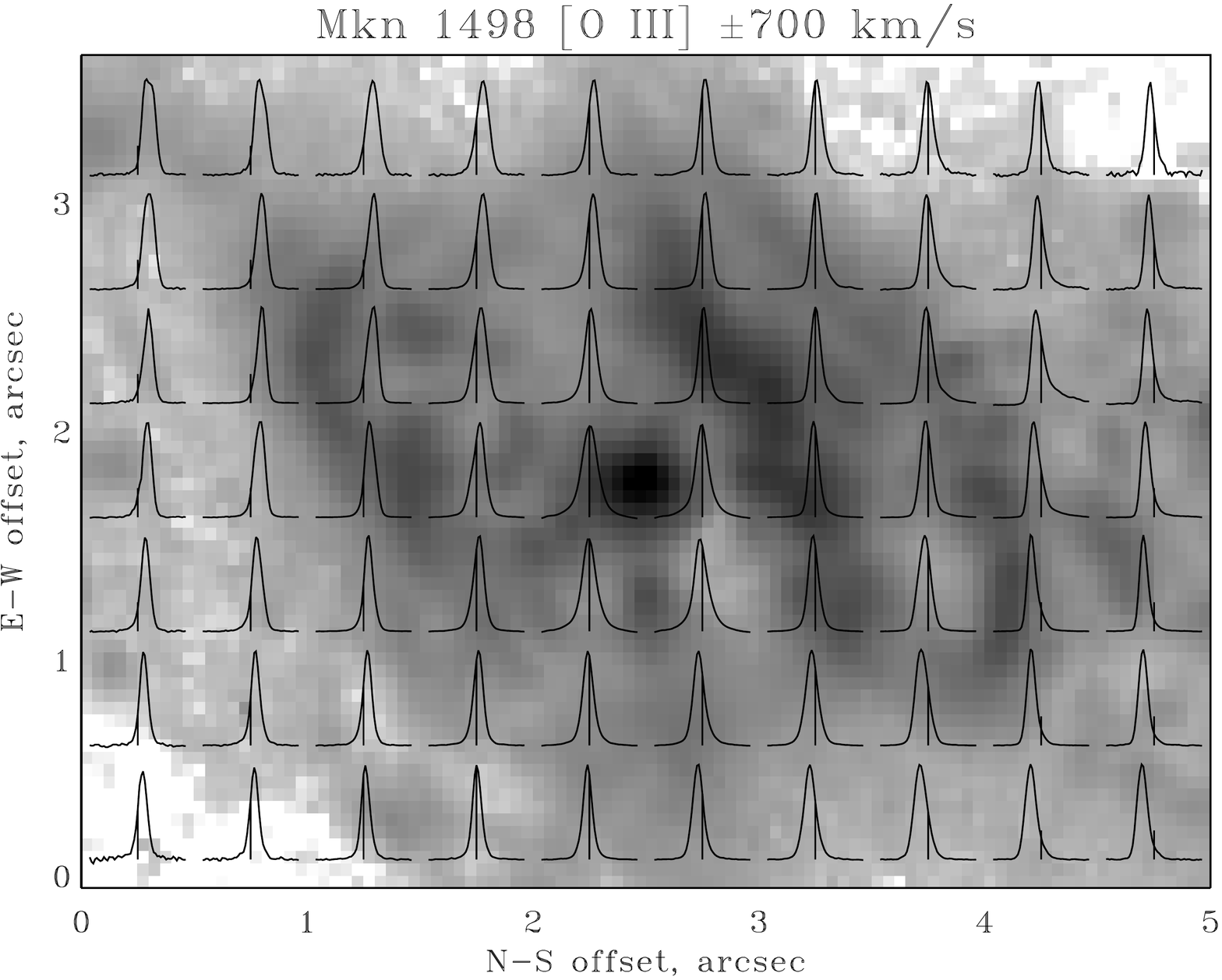} 
\caption{(continued)} 
\end{figure*} 

\addtocounter{figure}{-1}

\begin{figure*} 
\includegraphics[width=130.mm,angle=90]{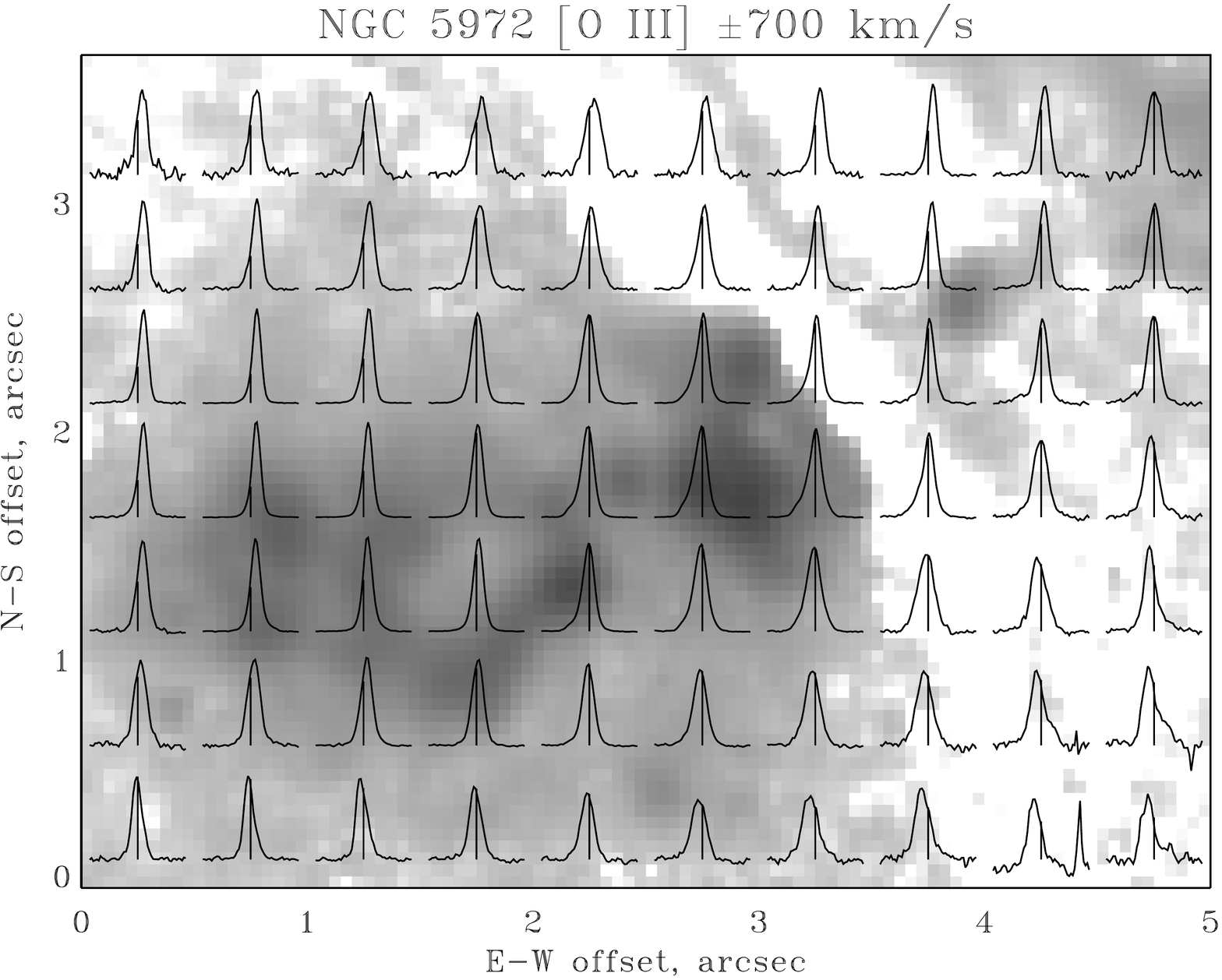} 
\caption{(continued)} 
\end{figure*} 

\addtocounter{figure}{-1}

\begin{figure*} 
\includegraphics[width=130.mm,angle=90]{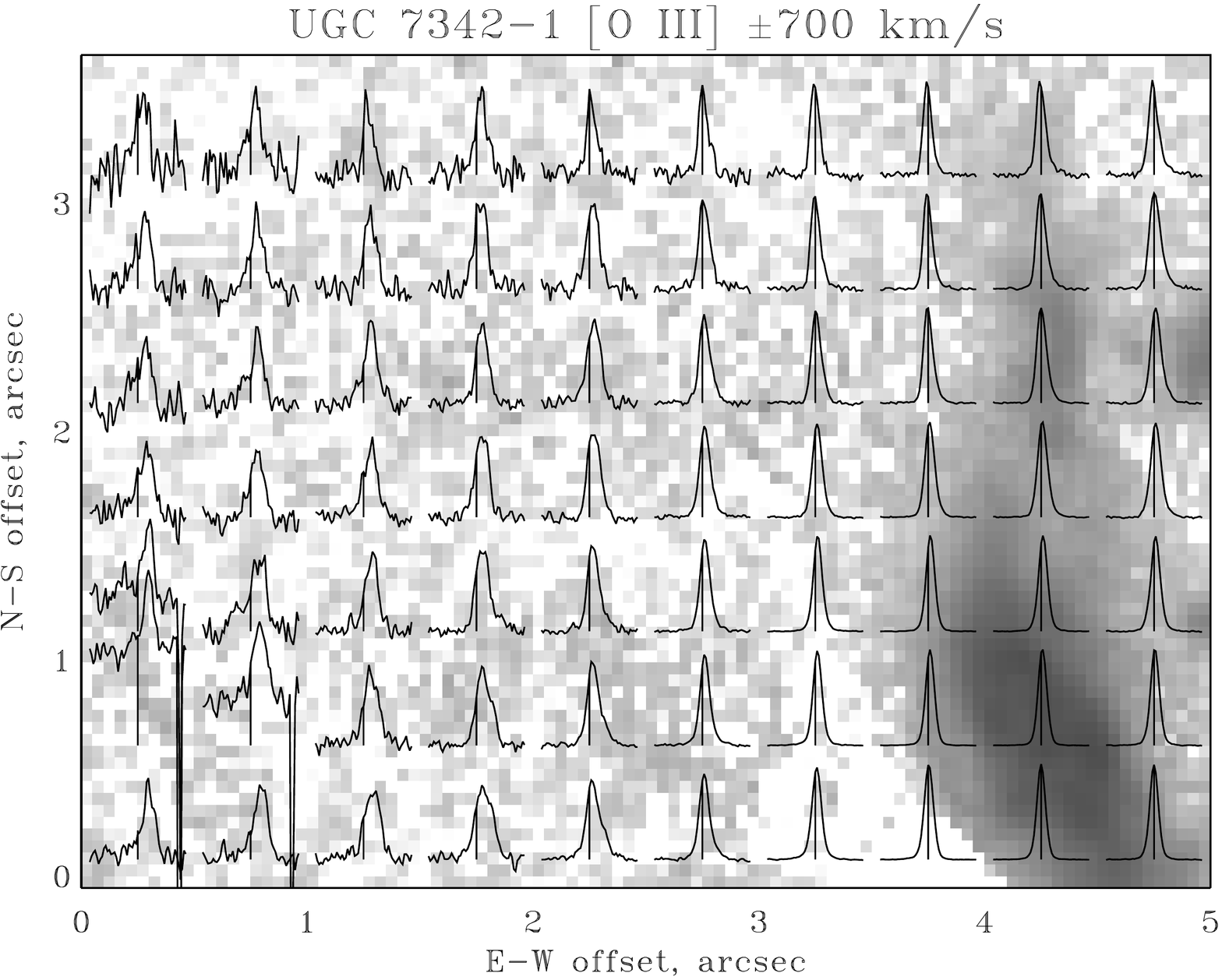} 
\caption{(continued)} 
\end{figure*} 

\addtocounter{figure}{-1}

\begin{figure*} 
\includegraphics[width=130.mm,angle=90]{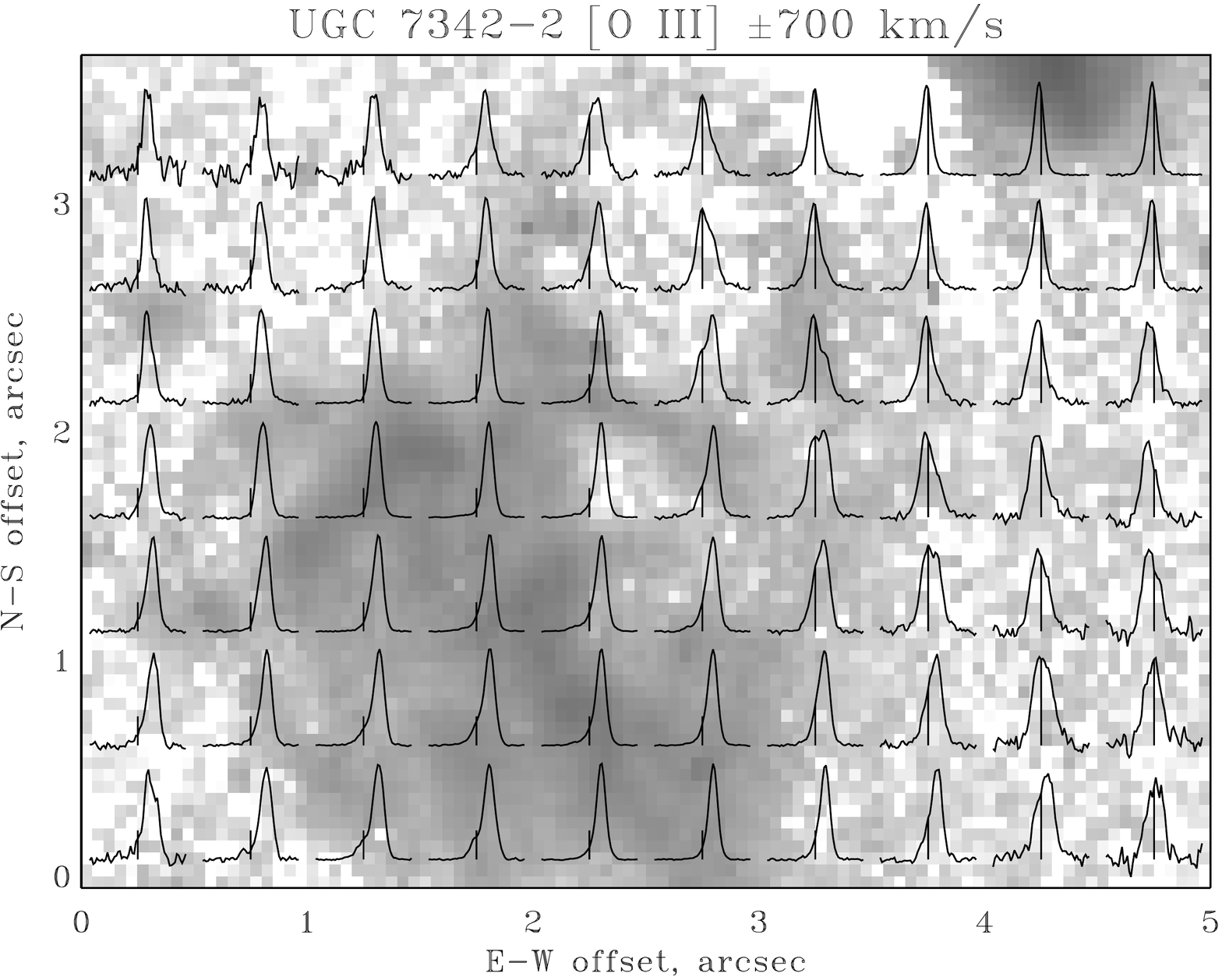} 
\caption{(continued)} 
\end{figure*}

In Mkn 1498 the ringlike emission features which are prominent at H$\alpha$ are 
dominated kinematically by rotation, possibly with one outflowing feature adjacent to to the nucleus 
(Fig. \ref{fig-mkn1498velfields}). The implied rotation axis is
close to the minor axis of the largest emission-line structures, making it almost parallel to
the large-scale radio structure. In the line profiles, red wings occur along the outer ring to the south,
and about 2\arcsec\ east of the nucleus along
the other ring (but not the rest of it). The inner arc structure to the northwest of the nucleus
has narrow line profiles, but departs from the kinematic pattern of the velocity peaks elsewhere in the ring. 

\begin{figure*} 
\includegraphics[width=148.mm,angle=0]{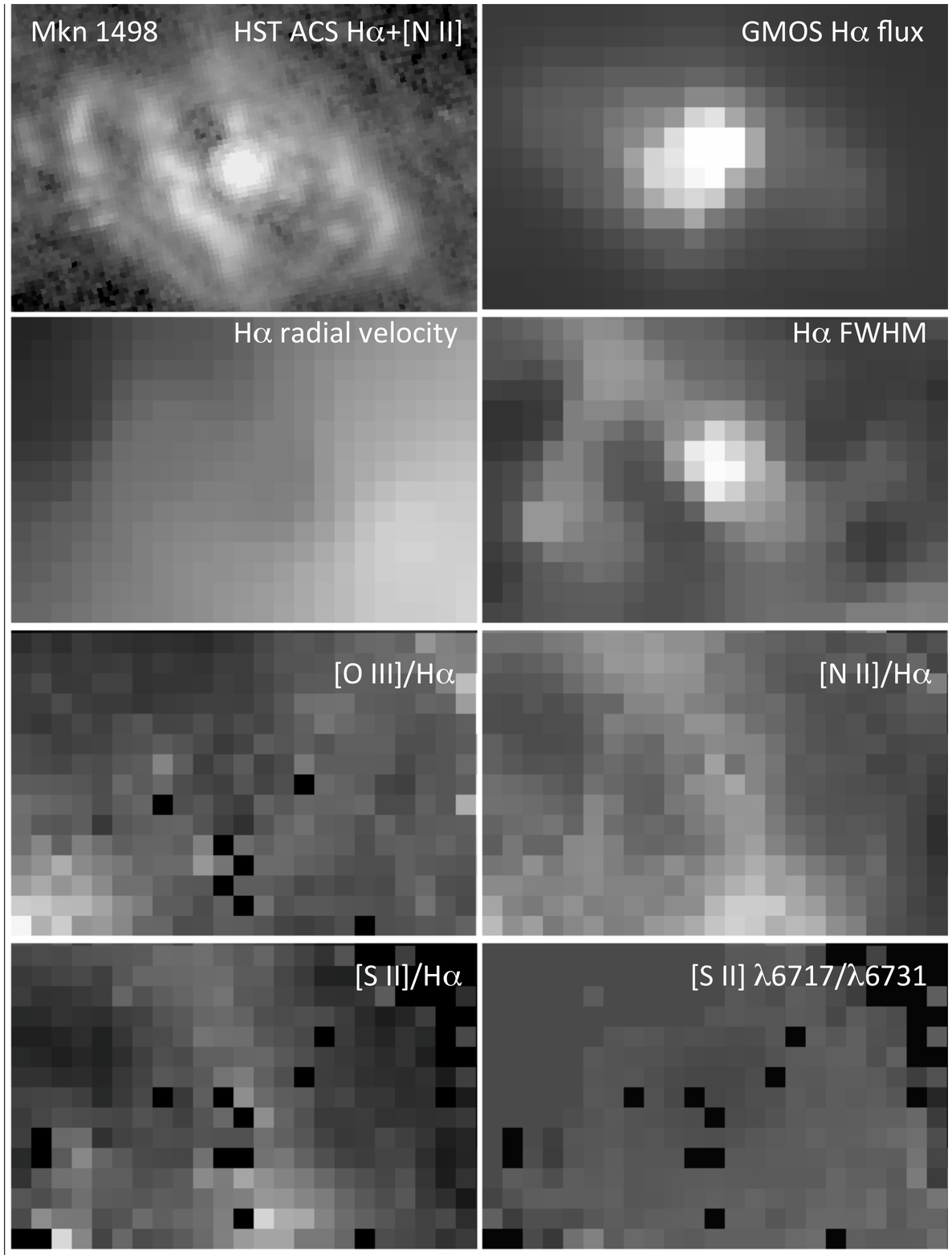} 
\caption{Summary of GMOS IFU results in the inner $3.4 \times 4.8$ arc seconds of Mkn 1498.
North is at the right, east at the top.
The {\it HST} narrowband image includes both H$\alpha$ and [N II] lines. For all quantities, lighter shades represent larger values.} 
\label{fig-mkn1498velfields} 
\end{figure*} 

As seen in Fig. \ref{fig-lineprofiles}, UGC 7342 shows substantial regions with double
emission-line peaks, likely evidence of outflow in at least one component. We examine 
this in more detail by 
fitting sets of 2 Gaussian [O III] profiles where this improves the fit, 
constraining them
to have equal widths for consistency across a range of signal-to-noise ratios.
Similar behavior is seen in H$\alpha$ and H$\beta$, but H$\beta$ is weaker
and H$\alpha$ is affected by overlap with [N II] components.
The results are shown in Fig. \ref{fig-ugc7342both}.
Only peak separations $> 100$ km s$^{-1}$ are detected, in view of the typical
line width of 230 km s$^{-1}$ FWHM. Component separations range up to 270 km s$^{-1}$.
The pattern of component velocities is not a simple two-sided outflow, and may have
more than one origin. The outer part of the filamentary emission region about 3.5\arcsec\ 
NE of the nucleus shows velocity splitting. This cloud hosts the most intense emission seen in the redder component (whose
radial velocity changes from place to place), while its outer part shows additional
blue wings (Fig. \ref{fig-lineprofiles}).
In addition, perpendicular to the main axis of the emission regions,
there are extensive regions with split line profiles, much of which is from emission too
spatially diffuse to be well detected in the HST images.

Using the BTA Fabry-Perot global velocity map from Paper 1 as a guide to the
overall velocity field, which is well fitted by a rotating and slightly warped disk,
we find that the redder component is the more quiescent one to the south of the
nucleus, while the bluer component fills this role to its north and northwest
(where large velocity splittings occur in diffuse gas). East of the nucleus,
the HST images do show a faint filament which could be associated with one of these
velocity structures. The areas with large local velocity dispersion $\sigma_v$ in the BTA data 
just to the
north of the nucleus are a good match to regions of line splitting in the GMOS
data cubes.

We are left with a complex empirical picture for kinematics in
the inner few kpc of UGC 7342. Emission-line components suggest outflow, or
at least motions departing from the overall rotation pattern by as much as
260 km s$^{-1}$, in regions of diffuse, low-surface-brightness emission
perpendicular to the main emission regions, with the strongest off-rotational
emission in the outer parts of a bright cloud, part of which also shows
blue wings beyond a simple 2-component line profile. We might speculate that
this could be entrainment of denser gas by a low-density outflow such as seen
west of the nucleus. Curiously, the brightest [O III] structure, resembling an
inclined ring just SW of the AGN position, has no distinct signature in radial
velocity.

\begin{figure*} 
\includegraphics[width=140.mm,angle=270]{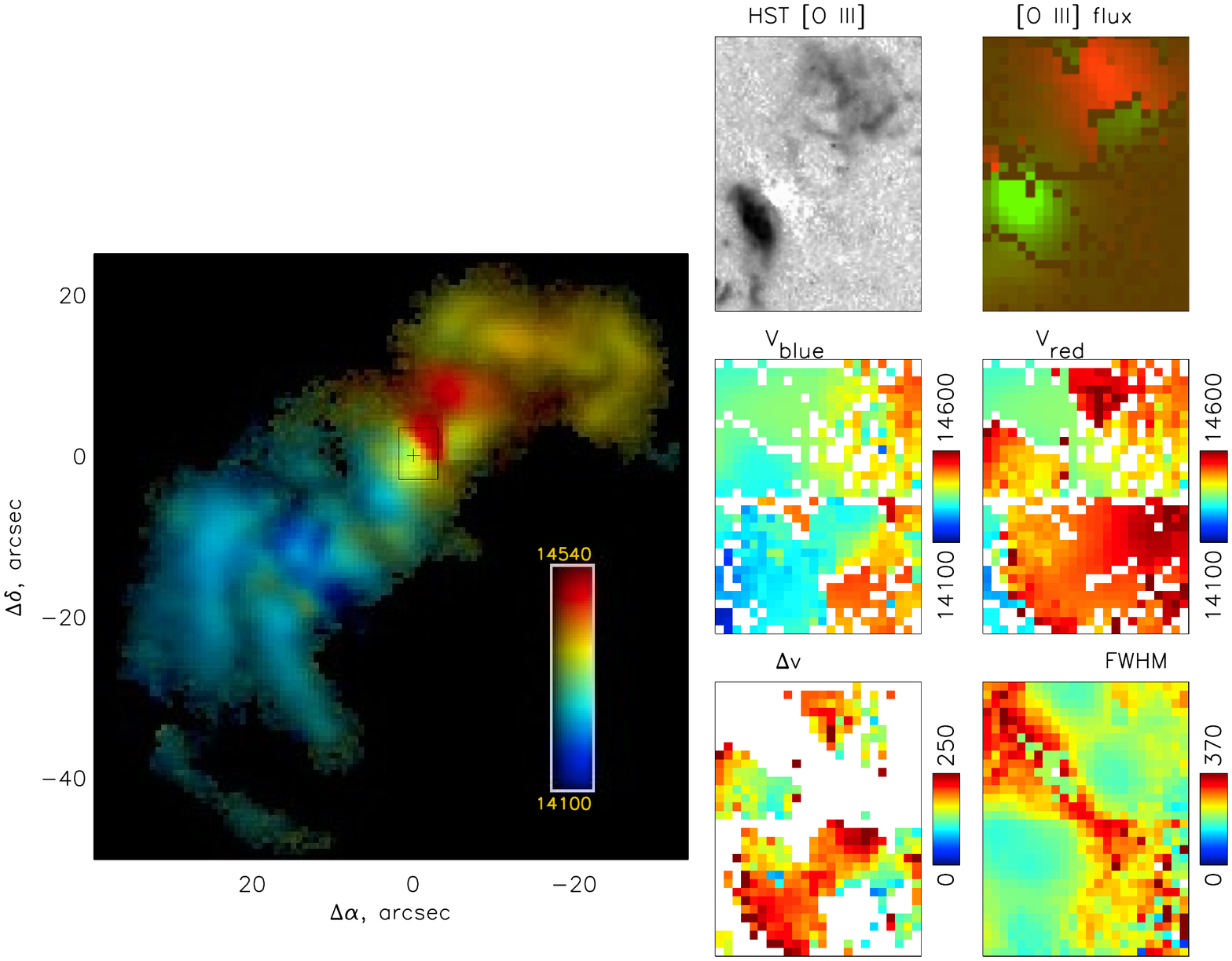} 
\caption{Results of 2-component Gaussian-blend fits to [O III] emission in UGC 7342.
The HST ACS [O III] image is scaled logarithmically to more fully display 
the structures near the core.
The two GMOS IFU fields are abutted at their nominal locations here, giving
contiguous coverage over $4.8 \times 6.4$\arcsec\ . North is at the top, east to
the left. 
The intensities of red and blue (shown in green for contrast) components are
scaled independently for better visibility. Some regions of large FWHM are adjacent 
to double-peaked
areas, presumably representing unresolved multiple components.
For comparison, the left panel shows the Fabry-Perot velocity field of the entire
galaxy as measured with the BTA \citep{paper1}, combining hue for radial velocity with
luminance for [O III] flux.} 
\label{fig-ugc7342both} 
\end{figure*}

The Teacup AGN was shown to have resolved outflow signatures by \cite{Harrison}. We
find  double or asymmetric profiles extending 2\arcsec\ north, 4\arcsec\ west, $>1.5$\arcsec\ 
south, and 2\arcsec\ east
in a roughly circular zone. What appears as an [O III] peak in the center of the loop on the underlying
HST image may be affected by residuals from cosmic-ray events.
The Gemini data are better at detecting emission which is diffuse and at low surface brightness, as in the center of the loop, and show no distinct emission-line component there.
Blue wings dominate east of the nucleus, red wings to the west; in many locations they extend to 1000 km s$^{-1}$ from the systemic velocity.


NGC 5972 shows significant velocity structure, and a nearly constant ionization level (Fig. \ref{fig-ngc5972ifu}). It shows double and asymmetric peaks near the dust lane, which can occur strictly from
obscuration in a continuous velocity field, and in the inner part of the emission-line
loop to its east. Overall, the emission-line loop to the east of the nucleus
shows no discernible velocity signature; it does have narrower line
profiles than its surroundings, perhaps because its stronger emission dominates the
more diffuse gas elsewhere along the line of sight. There is larger-scale velocity structure
in the GMOS field spanning $\approx 100$ km s$^{-1}$, not matching the loop morphology or
location, part of the departures from a fit to circular motions seen 
near the nucleus in the
wide-field BTA Fabry-Perot map \citep{paper1}. The [O III]/H$\alpha$ ratio declines with radius
from the nucleus at the same rate in the loop and its surroundings; the loop does not stand out in this
parameter. It has lower [N II]/H$\alpha$ and [S II]/H$\alpha$ ratios than its surroundings,
but not  [O III]/H$\alpha$ or density-sensitive  [S II] $\lambda 6717 / \lambda 6731$ ratio.
The loop region 2.0\arcsec\ E of the nucleus has [N II] $\lambda 6583$/H$\alpha$=0.33 (0.71 in the surroundings) and
[S II] $\lambda \lambda 6717+6731$/H$\alpha$=0.36 (0.70). Both line ratios are more sensitive
to abundances than ionization parameter under AGN-like photoionization, unlike [O III]/H$\alpha$
\citep{SB98}. Very broadly, the line ratios in this small-scale loop of emission show lower abundances than its surroundings,
which could suggest, contrary to its appearance, that it consists of infalling or recently accreted
gas. Together, these properties seem more like infall than an AGN-driven outflow; the alternative
would require a fairly contrived set of circumstances, such as motion in the plane of the sky and
the outflow itself being unseen (for example, too hot for optical emission) and entraining
material which originated in a low-luminosity companion disrupted by the obvious merger
history. Young stars might give the observed line ratios, in which case it would be a coincidence
that the [OIII]/H$\alpha$ ratio matches the surrounding AGN-ionized gas so closely, but 
no luminous star clusters appear in WFC3 continuum images \citep{paper1}.

\begin{figure*} 
\includegraphics[width=140.mm,angle=270]{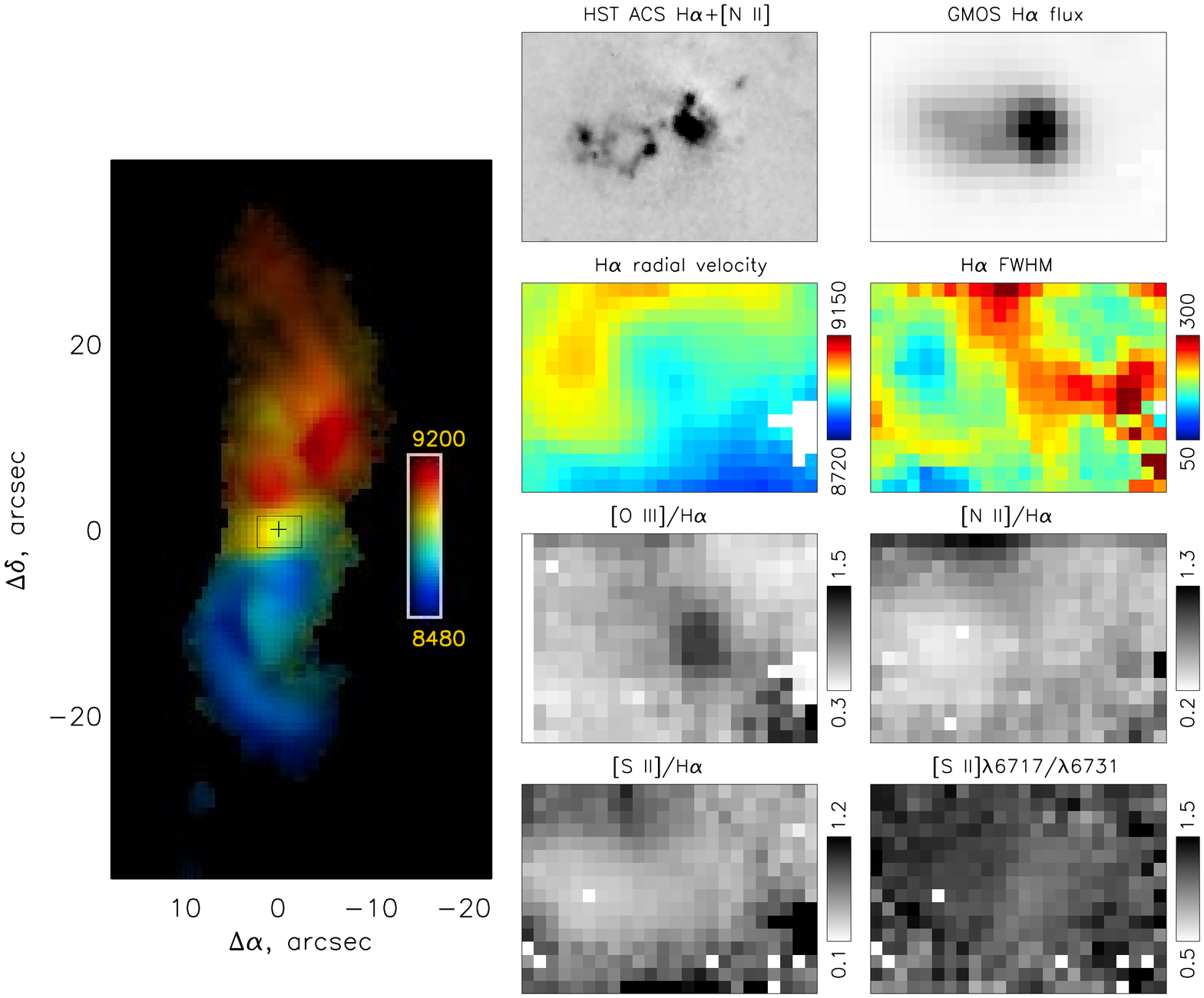} 
\caption{Summary of GMOS IFU results in the inner $3.4 \times 4.8$ \arcsec\  of NGC 5972.
North is at the top, east to the left.
The {\it HST} narrowband image includes both H$\alpha$ and [N II] lines. The peak
FWHM of 326 km s$^{-1}$ occurs 0.2\arcsec\ east of the peak H$\alpha$ flux.
For comparison, the left panel shows the Fabry-Perot velocity field of the entire
galaxy as measured with the BTA \citep{paper1}, combining hue for radial velocity with
luminance for [O III] flux.} 
\label{fig-ngc5972ifu} 
\end{figure*}


\section{Comparison: Luminosity Changes in Accretion Simulations}

We can compare the luminosity changes we measure to simulations such as \cite{Novak} to see how common such dramatic changes
are predicted to be for various feedback properties. Existing simulations have time resolution of the order of $10^4$ years, so there may be just enough information to tell what 
behavior is expected on timescales close to $10^5$ years. Fig. \ref{fig-simtimespan} shows
how common various levels of drops in accretion luminosity (derived from the simulations as 
$L/L_{Edd}$) are predicted to be when sampled on various timescales. We emulate the 
measurements by smoothing the starting luminosity at each timestep by 30\% of the timespan
being sampled, which mimics the effects of finite depth along the line of sight and increased
recombination timescale at the typically lower densities farther from the AGN. Although the
time resolution is not really adequate to test what happens when sampling a few $10^4$ years,
it is encouraging that this simulation shows a few per cent of its time steps which would
satisfy our selection criteria for AGN fading if suitable surrounding gas were present
(Table \ref{tbl-simtimespan}). The table lists the fraction of time steps with luminosity drops at
least as steeply as the listed ratios, evaluated for various values of $\Delta T$ according to
$L(T)/ \overline{L(T - \Delta T)}$ where the averaging in the denominator is over $0.3 \Delta T$ as above.
As \cite{Novak} note, their simulations show long periods of near-constant accretion, and
other periods showing many cycles of nearly periodic changes.

\begin{figure*} 
\includegraphics[width=148.mm,angle=0]{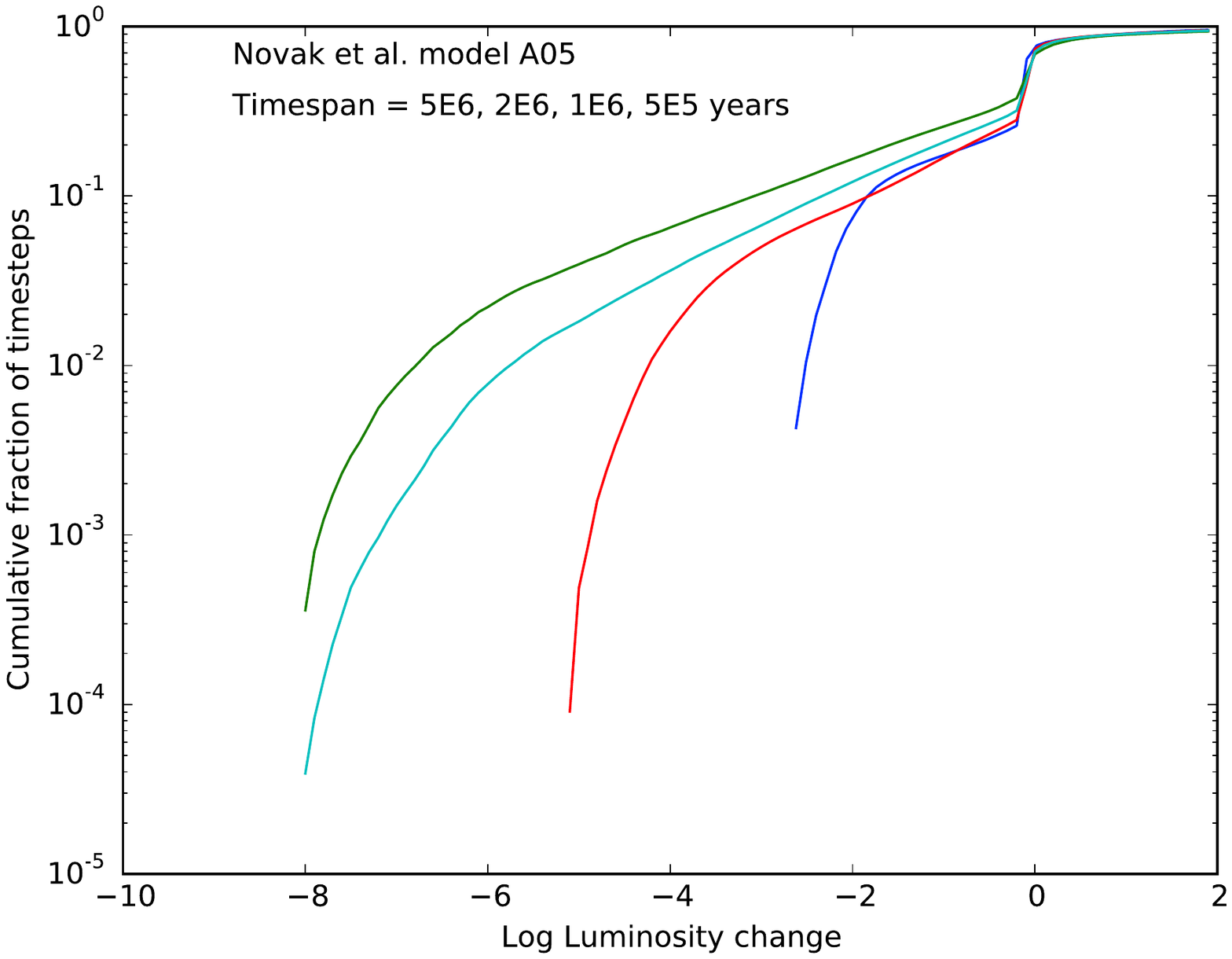} 
\caption{Fraction of timesteps from simulation A05 on \cite{Novak} as a function of luminosity
drop $L(T)/ \overline{L(T - \Delta T)}$, evaluated on timescales from $5 \times 10^5$ to $5 \times 10^6$ years. For each
time series, the starting luminosity was smoothed over 30\% of the relevant timespan to
roughly model the effects of finite cloud depth along the line of sight and longer recombination
ties at low densities. Lack of points to the lower right indicates time resolution effects; at
longer timescales we may be seeing effects of physically driven timescales in the
simulation.} 
\label{fig-simtimespan} 
\end{figure*} 

\begin{deluxetable}{lcccc}
\tablecaption{Simulated time fractions in low states \label{tbl-simtimespan}}
\tablewidth{0pt}
\tablehead{
\colhead{Log (luminosity drop)} & \colhead{Timespan $5 \times 10^6$ yr} & \colhead{Timespan $2 \times 10^6$ yr}  & \colhead{Timespan $10^6$ yr}& \colhead{Timespan $5 \times 10^5$ yr} }
\startdata
-1    &   0.26    &  0.20   & 0.16    &  0.16\\
-2     &  0.15    &  0.12   & 0.085   & 0.063\\
-3     & 0.098   & 0.067  & 0.050  &--- \\
-4     & 0.063   & 0.037  & 0.015   & --- \\
-5    & 0.039   & 0.019  & 0.0005 & ---  \\

\enddata
\tablecomments{Fraction of time steps in the simulation where the measured luminosity drop with respect to the averaged past value is more extreme than the reference value for each column.}
\end{deluxetable}

Some simulations indicate that smaller linear scales for changes in the inflow rate (from clumpy accretion or changes due to feedback from the AGN) yield faster changes in accretion rate
\citep{GaborBournaud}.

\section{Summary}
We present HST narrowband imaging and Gemini integral-field optical spectra for a 
set of fading AGN, selected for shortfalls in the energy budget between giant ionized clouds and the nuclei themselves suggesting significant reduction in ionizing photons within the last
$\approx 2 \times 10^4$ years. In particular, we use recombination balance to estimate the
history of radiative output, finding that the feature in common is a radial drop in luminosity
within 20,000 years before our direct view of the nucleus. For earlier times, we see a range
of behaviors - constant, brightening, fading - which are echoed in similarly reconstructed
histories of PG QSOs where [O III] HST imaging is available. 

In contrast to many of the extended emission clouds around radio-loud AGN, these are
rotationally dominated and show only very localized outflow velocities greater than
$\approx 100$ km s$^{-1}$. These EELRs are mostly externally illuminated tidal debris
rather than wind material \citep{paper1}.

As in \cite{mnras2012}, these results continue to support the idea that AGN with
extended emission regions are bright for periods of $10^4 - 10^5$ years at a time,
with substantially fainter episodes interspersed (which fits with the idea that the opposite
behavior - brightening -  can be seen in the number of X-ray bright AGN without substantial narrow-line regions; \citealt{Schawinski15}). Most galaxies do not have
large reservoirs of (extraplanar) cold gas to show such behavior, so we can trace it 
preferentially in interacting or merging systems where warped disk gas or tidal tails
provide a screen to be ionized by escaping ionizing photons. 

Among EELR hosts from the Galaxy Zoo sample, fading AGN make up $\approx 40$\% of the total, leading to the simple estimate of high-accretion timescales in \cite{mnras2012}.
This leaves open the possibility that fading cases could be a specific subset of the whole population. 
For example, inspiral of bound binary supermassive black holes could disturb
a pre-existing accretion disk, while significant changes in the direction of ionizing radiation could also result from the kinds of tilted disk discussed by 
\cite{LawrenceElvis}. In that case, shadowing by the disk itself would leave ionization ``cone"
shapes which are set by the intersection of two cones of different center directions
and opening angles.

\acknowledgments

This work was supported by NASA through STScI grants HST-GO-12525.01-A and -B.
Some of the data presented in this paper were obtained from the Mikulski Archive for Space Telescopes (MAST). 
This research has made use of NASA's Astrophysics Data System, and
the NASA/IPAC Extragalactic Database (NED) which is operated by the Jet Propulsion Laboratory, California Institute of Technology, under contract with the National Aeronautics and Space Administration. We thank Linda Dressel for advice on setting up the HST observations, especially
reducing intrusive reflections. Identification of this galaxy sample
was possible through the efforts of nearly 200 Galaxy Zoo volunteers; we are grateful for their contributions, and thank once more the
list of participants in \cite{mnras2012}. This work was partly supported by a grant from the 
President of the Russian Federation (MD-3623.2015.2).  The observations obtained with the 6-m telescope of the SAO of the RAS were carried out with the financial support of the Ministry of Education and Science of the Russian Federation (contracts no. 16.518.11.7073 and
14.518.11.7070). We benefited from conversations with Hai Fu, Larry Rudnick,
Kelly Holley-Bockelmann, Tamara Bogdanovic, and Stephanie Juneau.
Greg Novak kindly provided detailed results of his numerical simulations.
We thank Dara Norman and Kathy Roth for key help in setting up the Gemini
observations, James Turner for sharing experience in subtleties of their
processing, and Gemini staff observers M. Hoenig, A.-N. Chene, J. Chavez, M. Pohlen,
L. Fuhrman, and J. Ball for obtaining these data in queue mode.

C. J. Lintott
acknowledges funding from The Leverhulme Trust and the STFC Science in Society Program. 
W.P. Maksym is grateful for support by the University of Alabama Research Stimulation Program.
V.N. Bennert acknowledges assistance from a National Science Foundation (NSF) Research at Undergraduate Institutions (RUI) grant AST-1312296. Note that findings and conclusions do not necessarily represent views of the NSF. A. Moiseev is also grateful for
the financial support of the ``Dynasty" Foundation. 
K. Schawinski was supported  by a NASA Einstein Fellowship at Yale, and gratefully acknowledges support
from Swiss National Science Foundation Grant PP00P2 138979/1. Galaxy Zoo was made
possible by funding from the Jim Gray Research Fund from Microsoft and The Leverhulme Trust.

This publication makes use of data products from the Wide-field Infrared Survey Explorer, which is a joint project of the University of California, Los Angeles, and the Jet Propulsion Laboratory/California Institute of Technology, funded by the National Aeronautics and Space Administration.

{\it Facilities:} 
\facility{HST (ACS, WFC3, WFPC2)},
\facility{BTA}, \facility{Gemini Gillette}
\facility{WISE}




\clearpage

\end{document}